\documentclass[11pt,draftcls,onecolumn]{IEEEtran}
\usepackage{cite}
\usepackage{amsmath,amssymb,amsfonts}
\usepackage{algorithmic}
\usepackage{graphicx}
\usepackage{textcomp}
\usepackage{wrapfig}

\usepackage{amsfonts,bm,amsmath,comment,color,amssymb,subfigure,times}
\usepackage[]{graphicx}
\usepackage[linesnumbered,ruled]{algorithm2e}
\usepackage{multirow}
\usepackage{makecell}
\usepackage{epstopdf}
\usepackage{caption}
\def\BibTeX{{\rm B\kern-.05em{\sc i\kern-.025em b}\kern-.08em
    T\kern-.1667em\lower.7ex\hbox{E}\kern-.125emX}}

\def\x{{\mathbf x}}

\def\bLam{{\mathbf{\Lambda}}}

\def\bzero{{\mathbf{0}}}

\def\c1{{\textcircled{a}}}
\def\ba{{\mathbf{a}}}
\def\bb{{\mathbf{b}}}

\def\bd{{\mathbf{d}}}

\def\bg{{\mathbf{g}}}
\def\bh{{\mathbf{h}}}

\def\bm{{\mathbf{m}}}
\def\bn{{\mathbf{n}}}

\def\bp{{\mathbf{p}}}

\def\br{{\mathbf{r}}}
\def\bs{{\mathbf{s}}}

\def\bu{{\mathbf{u}}}

\def\bx{{\mathbf{x}}}
\def\by{{\mathbf{y}}}
\def\bz{{\mathbf{z}}}
\def\bA{{\mathbf{A}}}
\def\bB{{\mathbf{B}}}
\def\bC{{\mathbf{C}}}
\def\bD{{\mathbf{D}}}
\def\bE{{\mathbf{E}}}

\def\bG{{\mathbf{G}}}
\def\bH{{\mathbf{H}}}
\def\bI{{\mathbf{I}}}
\def\bJ{{\mathbf{J}}}

\def\bN{{\mathbf{N}}}
\def\bM{{\mathbf{M}}}

\def\bP{{\mathbf{P}}}
\def\bQ{{\mathbf{Q}}}
\def\bR{{\mathbf{R}}}
\def\bS{{\mathbf{S}}}
\def\bT{{\mathbf{T}}}
\def\bU{{\mathbf{U}}}
\def\bV{{\mathbf{V}}}

\def\bX{{\mathbf{X}}}
\def\bY{{\mathbf{Y}}}
\def\bZ{{\mathbf{Z}}}

\def\bzero{{\mathbf{0}}}

\def\tr{{\textrm{{tr}}}}
\def\T{{\top}}
\renewcommand\Re{{\textrm{Re}}}
\renewcommand\vec{{\textrm{vec}}}
\def\HH{{\dagger}}
\def\blambda{{\boldsymbol{\lambda}}}

\begin{document}
\title{Relative Entropy-Based Waveform Optimization for Rician Target Detection with Dual-Function Radar Communication Systems}
\author{Xuyang Wang, Bo Tang, Wenjun Wu and Da Li
\thanks{This work was supported in part by the National Natural Science Foundation of China under Grant 62171450 and 61671453, the Anhui Provincial Natural Science Foundation under Grant 2108085J30. (Corresponding author: Bo Tang.)}
\thanks{The authors are with the College of Electronic Engineering, National University of Defense Technology, Hefei 230037, China. (e-mail: wangxuyang@nudt.edu.cn; tangbo06@gmail.com; wuwenjun.1010@nudt.edu.cn; lida@nudt.edu.cn).}
\thanks{}}

\markboth{Preprint}
{Shell \MakeLowercase{\textit{et al.}}: A Sample Article Using IEEEtran.cls for IEEE Journals}
\maketitle
\begin{abstract}
In this paper, we consider waveform design for  dual-function radar-communication systems based on multiple-input-multiple-out arrays. To achieve better Rician target detection performance, we use the relative entropy associated with the formulated detection problem as the design metric. We also impose a multi-user interference energy constraint on the waveforms to ensure the achievable sum-rate of the communications. Two algorithms are presented to tackle the nonlinear non-convex waveform design problem. In the first algorithm, we derive a quadratic function to minorize the objective function. To tackle the quadratically constrained quadratic programming problem at each iteration, a semidefinite relaxation approach followed by a rank-one decomposition procedure and an efficient alternating direction method of multipliers (ADMM) are proposed, respectively. In the second algorithm, we present a novel ADMM algorithm to tackle the optimization problem and employ an efficient minorization-maximization approach in the inner loop of the ADMM algorithm. Numerical results demonstrate the superiority of both algorithms. Moreover, the presented algorithms can be extended to synthesize peak-to-average-power ratio constrained waveforms, which allows the radio frequency amplifier to operate at an increased efficiency.
\end{abstract}

\begin{IEEEkeywords}
DFRC system, MIMO array, waveform design, Rician target, relative entropy.
\end{IEEEkeywords}

\section{Introduction}
\label{sec:introduction}
\IEEEPARstart{R}{adar} and communications are two important applications of radio technology. Traditionally, the two kinds of systems are allocated to different frequency bands and work separately. However, with the rapid technological progress, such as the rapid deployment of the fifth-generation (5G) communications and the internet of things, the number of communication systems grows quickly and the demands to access to a wider spectrum by these systems are ever-increasing. As a consequence, the possible spectral overlap between communication and radar systems can lead to severe mutual interferences \cite{Griffiths2013Challenge,Griffiths2015Spectrum}. To reduce the mutual interference and improve the compatibility in spectrally crowded environments, several schemes have been proposed. One scheme to enhance the spectral coexistence between radar and communication systems is by designing spectrally constrained waveforms. In \cite{Rowe2014SHAPE,Tang2019shape}, the authors presented several approaches to synthesize waveforms with desired spectral shapes. In \cite{Aubry2016Optimization,Aubry2016Multiple,Aubry2020MultiSpectrally,Tang2018ADMM}, the authors focused on the maximization of the signal-to-interference-plus-noise-ratio (SINR) under one or multiple spectral constraints.  In addition to designing spectrally constrained waveforms, the authors in \cite{Li2017Joint,Qian2018Transmit,Qian2018Joint} studied the co-design of radar and communication systems. By jointly optimizing the radar waveforms and the covariance matrix of the communication signals, the proposed methods therein can effectively alleviate the mutual interference between the two kinds of systems.

Recently, there have been growing interests in developing dual-function radar-communication (DFRC) systems to improve the spectral compatibility (see, e.g., \cite{Tavik2005RF,Liu2020JRC,Zhang2021Overview,Tang2022MFRF,Zhang2020Circulating,Sanson2021Cooperative,Bekar2021PSK-LFM,Liu2022PAPR}). A DFRC system can support target acquisition and data delivery simultaneously based on a shared  array.  Compared with other schemes, DFRC systems achieve an increased spectral efficiency, a reduced size, and a lower power consumption. The key of realizing a DFRC system is designing the waveforms properly. Currently,  the transmit waveforms of DFRC systems can be classified into three categories: communication-centric waveforms, radar-centric waveforms, and integrated waveforms. Communication-centric waveforms are mainly communication signals (e.g., orthogonal frequency division multiplexing (OFDM) signals \cite{Liu2019Robust,Liu2020OFDM}), which also serve for radar target detection and parameter estimation. Different from the communication-centric waveforms, radar-centric waveforms modulate the existing radar waveforms to deliver the information bits (e.g., in \cite{Liu2015MSK}, the authors used the minimum shift keying (MSK) signals to modulate the linear frequency modulated (LFM) waveforms).
Despite of the simplicity, communication-centric waveforms and radar-centric waveforms might suffer from the problems of envelope variations, high autocorrelation sidelobes, or low data rates. Inspired by the advantages provided by multiple-input-multiple-out (MIMO) systems \cite{LiMIMObook2008,Shi2020Spectrally,Sayin2020Design,Nan2020Beampattern,Cong2021Robust}, designing integrated waveforms for MIMO array-based DFRC systems (i.e., MIMO DFRC systems) have received considerable attentions.

In \cite{Liu2018Toward,Tang2020Dualfunction}, the authors proposed waveform design methods for MIMO DFRC systems. The aim was to minimize the multi-user interference (MUI) energy for communications and match a desired radar beampattern. In \cite{Tsinos2021Joint,Wu2022dual-function}, the authors proposed an algorithm to jointly design the transmit waveforms and receive filters for MIMO DFRC systems in the presence of clutter. In \cite{Liu2022Cramer}, the Cramer-Rao Bound (CRB) of radar parameter estimation was used as the metric to design waveforms for DFRC systems. Meanwhile, a constraint is enforced to guarantee the SINR of communication users. In \cite{Liu2020Multiuser}, the authors focused on the joint design of MIMO DFRC systems to minimize the synthesis error of the radar transmit beampattern and guarantee the performance of each communication user. In \cite{Liu2022Covariance}, the authors designed waveforms for DFRC systems based on maximizing the SINR of communication users under a constraint on the radar performance. In \cite{Yuan2021Spatio-Temporal}, the authors designed integrated waveforms based on information theory. The mutual information (MI) of the radar and communication systems were derived, respectively. Then a weighted sum of the two MIs was utilized to provide a tradeoff between the radar and communication performance.

In this paper, we design waveforms for target detection with a MIMO DFRC system. Note that such a problem is also considered in \cite{Tsinos2021Joint}, where a deterministic target model was assumed and the maximization of SINR was used as the design metric. Different from \cite{Tsinos2021Joint}, we describe the target response with the Rician model. Such an assumption is justified if there are multiple propagation paths between the DFRC system and the target.  To enhance the detection performance of the Rician target and for mathematical tractability, we utilize the relative entropy associated with the detection problem as the waveform design metric. In addition, an MUI energy constraint is enforced on the waveforms to guarantee the quality of communication service. To tackle the non-convex optimization problem, two methods are presented. In the first method, we leverage the minorization-maximization (MM) approach and minorize the objective function by a simpler quadratic function. At each iteration of the MM method, the quadratically constrained quadratic programming (QCQP) problem is tackled via a semidefinite relaxation (SDR) approach followed by a rank-one decomposition procedure and an alternating direction method of multipliers (ADMM), respectively. In the second method, we propose a novel ADMM algorithm to directly deal with the non-convex optimization problem and employ an efficient MM approach to tackle the sub-problem at each iteration. Moreover, to avoid the use of expensive linear amplifiers, we extend the presented methods to design peak-to-average-power ratio (PAPR) constrained waveforms. Simulations results demonstrate that when the quality of communication service is ensured, the waveforms synthesized by the presented algorithms can obtain better Rician target detection performance.

We organize the rest of this paper as follows. Section \ref{Sec:ProblemFormulation} derives the radar model and the communication model. Then the waveform design problem is formulated. In Section \ref{Sec:AlgorithmDesignMM}, an efficient MM approach is presented to tackle the waveform design problem.  In Section \ref{Sec:AlgorithmDesignADMM}, a novel ADMM algorithm is proposed. In Section \ref{Sec:NumericalExamples}, the effectiveness of the proposed algorithms are demonstrated via simulations. In Section \ref{Sec:conclusion}, we conclude the paper.

The notations used in the paper are displayed in Table I.
\begin{table}[!htbp] \label{tab:1}
\caption{List of Notations}
\begin{center}
\begin{tabular}{cl}
\hline
Symbol & Meaning \\
\hline
$\bA$ & Matrix \\
$\ba$ & Vector \\
$(\cdot)^{\T}$, $(\cdot)^{\HH}$ & Transpose and conjugate transpose\\
$\mathbb{C}$ & The domain of complex numbers\\
$\mathbb{R}$ & The domain of real numbers\\
$\bI$ & Identity matrix\\
$\textrm{blkDiag}(\bA;\bB)$ & A block diagonal matrix formed by $\bA$ and $\bB$ \\
$\bzero$ & Null matrix\\
$\textrm{tr}(\cdot)$ & The trace of a matrix\\
$\det(\cdot)$ & The determinant of a matrix\\
$\|\cdot\|_{2}$ & The Euclidean norm of a vector \\
$\|\cdot\|_{\textrm{F}}$ & The Frobenius norm of a matrix \\
$\bA\otimes \bB$ & Kronecker product of $\bA$ and $\bB$\\
vec($\bX$) & Column-wise stacking of $\bX$\\
mod($x,y$) & The remainder after division $x$ by $y$\\
$\lceil x\rceil$ & The nearest integer greater than or equal to $x$\\
Re$(\cdot)$ & The real-part of a complex-valued number\\
$\bA\succ\mathbf{0}$ $(\bA\succeq\mathbf{0})$& $\bA$ is positive definite (semi-definite)\\
\hline
\end{tabular}
\end{center}
\end{table}
\section{Signal Model and Problem Formulation} \label{Sec:ProblemFormulation}
\begin{figure}[!htbp]
\centering
{{{\includegraphics[width = 0.4\textwidth]{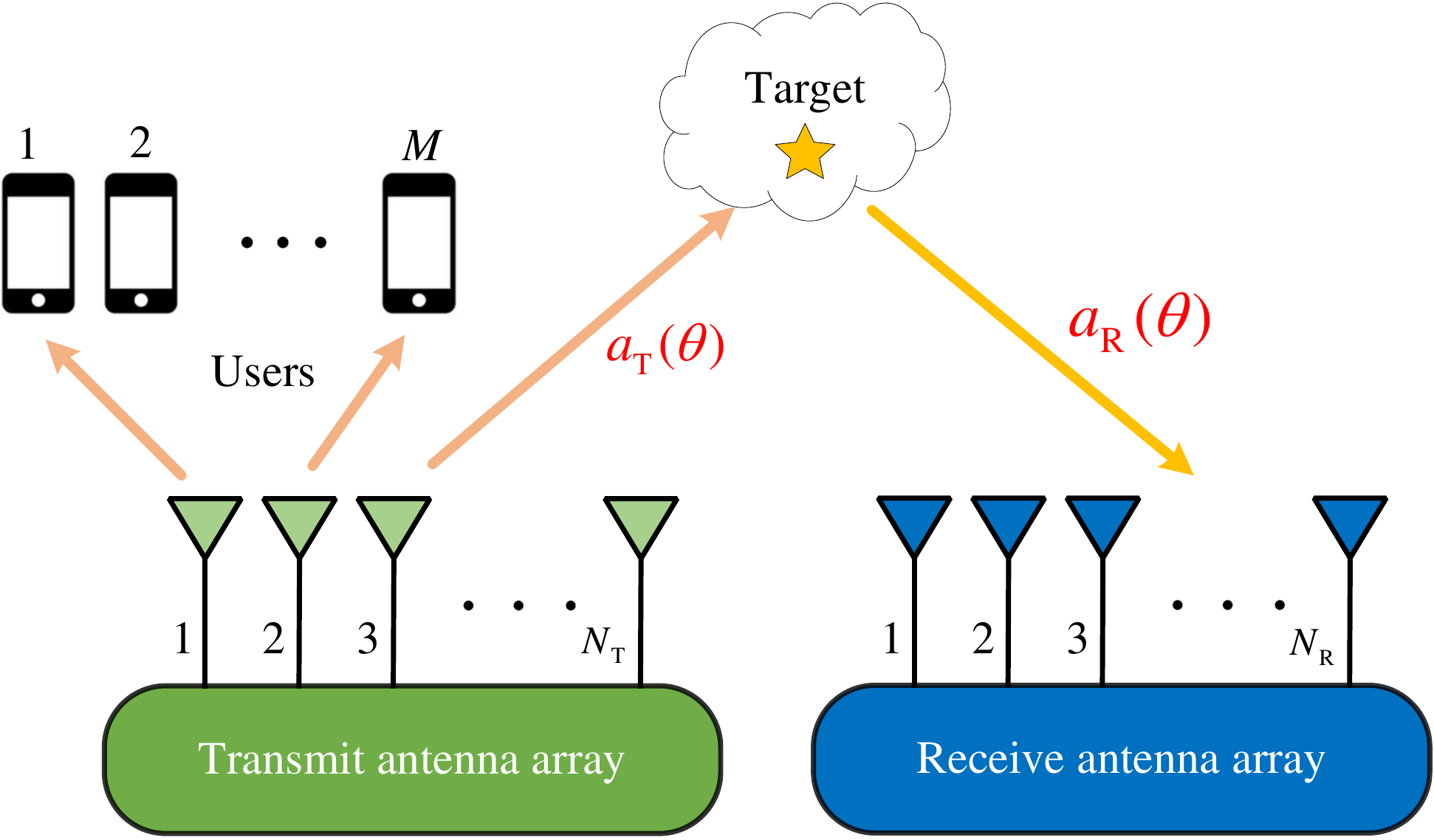}} \label{Fig:1}} }
\caption{The dual-function radar-communication system. }
\label{fig:1}
\end{figure}

As Fig. \ref{fig:1} shows, we consider a DFRC system based on a MIMO array. The MIMO array consists of $N_\textrm{T}$ transmit antennas and $N_\textrm{R}$ receive antennas. We denote the discrete baseband waveform of the $n$-th transmit antenna by $\bx_n \in \mathbb{C}^{L\times 1}$, $n=1,2, \cdots, N_\textrm{T}$, and let $\bX=[\bx_1,\bx_2,\ldots,\bx_{N_\textrm{T}}]\in \mathbb{C}^{L\times N_\textrm{T}}$ be the transmit waveform matrix, where $L$ is the code length. We assume the possible presence of one target and $M$ communication users.
To detect the target and communicate with multiple communication users simultaneously, next we establish the radar detection and multi-user communication models in the follows.

\subsection{Radar Detection Model}
Let $\bG\in \mathbb{C}^{N_\textrm{T}\times N_\textrm{R}}$ be the target response matrix.
Then the received signal due to the target reflections is given by
\begin{equation} \label{eq:SigModel}
  \bY = \bX\bG + \bN,
\end{equation}
where $\bN \in \mathbb{C}^{L\times N_\textrm{R}}$ denotes the receiver noise.
Let $\by = \vec(\bY)$, $\bg = \vec(\bG)$, and $\bn = \vec(\bN)$, then the signal model can be rewritten as
\begin{equation} \label{eq:SigMode2}
\by= \widetilde{\bX}\bg+\bn,
\end{equation}
where $\widetilde{\bX} = \bI_{N_\textrm{R}}\otimes \bX$. It is worth noting that the signal model in \eqref{eq:SigMode2} can be used for several kinds of MIMO systems (see \cite{Tang2019Spectrally,Tang2021RangeSpread,Wang2022Rician} for similar models), e.g., if we consider a coherent MIMO radar, the target response is modeled by
\begin{equation}
\bg = \sum_{k=1}^K \alpha_{t,k} \bb(\theta_{t,k}) \otimes \ba(\theta_{t,k}),
\end{equation}
where $\alpha_{t,k}$, $\theta_{t,k}$, $\ba(\theta_{t,k})$, and $\bb(\theta_{t,k})$ are the amplitude, the direction of arrival (DOA), the transmit steering vector, and the receive steering vector of the $k$th  scatterer ($k=1, \cdots, K$), respectively, and $K$ is the number of scatterers.

To determine the target presence, we establish the following hypothesis test:
\begin{equation}  \label{eq:hypothesistest}
\left\{
\begin{aligned}
\mathcal{H}_0:& \by=\bn, \\
\mathcal{H}_1:& \by= \widetilde{\bX}\bg+\bn.
\end{aligned}
\right.
\end{equation}
For target detection in the presence of multipath or knowledge mismatch, we can utilize a Rician model to describe the target response $\bg$. Indeed, compared with the Rayleigh model and the deterministic model, the Rician model is more general \footnote{Indeed, the Rayleigh model and the deterministic model can be treated as special cases of the Rician model \cite{Tang2016Rician,Wang2022Rician}.}. For a Rician target, the target response $\bg$ can be modeled by $\bg\sim\mathcal{CN}(\bg_d,\bR_\textrm{G})$, i.e., $\bg$ obeys a circularly-symmetric Gaussian distribution with mean $\bg_d$ and covariance matrix $\bR_\textrm{G}$. That is to say, the PDF of $\bg$ can be written as \cite{Kaystatistical1998}
\begin{equation} \label{eq:hypothesis}
  P(\bg) = \frac{1}{\pi^{N_\textrm{TR}}\det(\bR_\textrm{G})} \exp\left(-(\bg-\bg_d)^\HH \bR_\textrm{G}^{-1} (\bg-\bg_d)\right),
\end{equation}
where $N_\textrm{TR} = N_\textrm{T}N_\textrm{R}$.
Assume that the receiver noise is white Gaussian, with noise power of $\sigma^2$. It can be checked that the PDF of the observations under the two hypotheses, denoted by $P_0(\by)$ and $P_1(\by)$, can be expressed as follows:
\begin{align}\label{eq:hypothesis2}
  P_0(\by) &= \frac{1}{\pi^{N_\textrm{RL}}\sigma^{2N_\textrm{RL}}} \exp\left(-\sigma^{-2}\by^\HH \by\right) \nonumber \\
  P_1(\by) &=\frac{1}{\pi^{N_\textrm{RL}}\det(\bR_1)} \exp\left(-(\by-\widetilde{\bX}\bg_d)^\HH \bR_1^{-1} (\by-\widetilde{\bX}\bg_d)\right)
\end{align}
where $\bR_1 = \widetilde{\bX}\bR_\textrm{G}\widetilde{\bX}^\HH+\sigma^2\bI_{N_\textrm{RL}}$, and $N_\textrm{RL} = N_\textrm{R} L$.

We can derive the Neyman-Pearson (NP) detector based on the above assumptions  and analyze the associated detection probability \cite{Naghsh2012}. However, since the expression of the detection probability is too complicated, it is intractable to use the detection probability as the design metric. Alternatively, we resort to the relative entropy between the observations under the two hypotheses as the criterion (for radar waveform design based on relative entropy,  see, e.g.,  \cite{Naghsh2013Unified,Naghsh2017Information,Tang2010mimo}). Indeed, Stein's lemma demonstrates that the detection probability increases asymptotically with the relative entropy \cite{CoverInformationTheory1991}. Specifically, let $\textrm{D}(P_0||P_1)$ be the relative entropy between $P_0{(\by})$ and $P_1{(\by})$. Then for a fixed probability of false alarm ($P_{f_a}$), the probability  of detection ($P_d$) is related to the relative entropy by the following expression:
\begin{equation}  \label{eq:D}
\textrm{D}(P_0||P_1)=\lim\limits_{N\rightarrow \infty}\left(-\frac{1}{N}\log(1-P_d)\right),
\end{equation}
where $N$ is the number of independent and identically distributed (i.i.d.) samples. Therefore, we use $\textrm{D}(P_0||P_1)$ as the waveform design metric to obtain better target detection performance.
By using the assumption in \eqref{eq:hypothesis2}, the relative entropy between $P_0(\by)$ and $P_1(\by)$ is given by
\begin{align} \label{eq:relativeentropy}
\textrm{D}(P_0||P_1)
=&\int P_0(\by)\log\frac{P_0{(\by})}{P_1{(\by})}d\by \nonumber \\
=&\log\det(\bR_1) + \tr[\bR_1^{-1}(\widetilde{\bX}\bg_d\bg_d^\HH\widetilde{\bX}^\HH+\sigma^2\bI_{N_\textrm{RL}})] \nonumber \\
&-N_\textrm{RL}(1+\log\sigma^2).
\end{align}
\subsection{Multi-user Communication Model}
The communication signal received by the $M$ users is given by
\begin{equation} \label{eq:SigModelc}
  \bY_\textrm{c} = \bH\bX^\T + \bZ,
\end{equation}
where $\bH=[\bh_1,\bh_2,\ldots,\bh_{M}]^\T\in \mathbb{C}^{M\times N_\textrm{T}}$ denotes the channel matrix, $\bh_m$ represents the channel vector of the $m$th user ($m=1,2,\cdots,M$), and $\bZ$ denotes the receiver noise at the communication receivers.
Let $\bS=[\bs_1,\bs_2,\ldots,\bs_{M}]^\T\in \mathbb{C}^{M\times L}$ be the desired signals at these communication receivers. Note that $\bY_\textrm{c}$ can be rewritten as \begin{equation} \label{eq:SigModelcc}
  \bY_\textrm{c} = \bS+\underbrace{\bH\bX^\T-\bS}_\textrm{MUI} + \bZ,
\end{equation}
where $\bH\bX^\T-\bS$ is called the MUI \cite{Mohammed2013Per-Antenna}. The MUI energy is defined by
\begin{equation} \label{eq:MUI}
  f(\bX) = \|\bH\bX^\T-\bS\|^2_\textrm{F}.
\end{equation}
In \cite{Mohammed2013Per-Antenna}, the authors indicated that the MUI energy is closely related to the achievable sum-rate for the multiple users. A smaller MUI energy results in a larger SINR  and a higher achievable sum-rate. Therefore, we can constrain the MUI energy to be lower than a small value to guarantee the multi-user communication performance.

\subsection{Problem Formulation}
To improve the target detection performance and guarantee the achievable sum-rate for communications, we use the relative entropy associated with the detection problem as the metric, and enforce an MUI energy constraint. The corresponding waveform design problem is formulated by
\begin{align} \label{eq:RobustDesign}
\max\limits_{\bX} & \ \textrm{D}(P_0||P_1)\nonumber\\
\textrm{s.t.} &\ f(\bX)\leq \epsilon,\nonumber\\
&\ \bX\in \mathcal{X},
\end{align}
where $\mathcal{X}$ is the waveform constraint set, and $\epsilon$ denotes the maximum allowed MUI energy. Due to the limited transmit energy of the waveforms, we enforce an energy constraint on the radar waveforms:
\begin{align*}
\textrm{tr}(\bX\bX^\HH)\leq P_\textrm{t},
\end{align*}
where $P_\textrm{t}$ is the transmit energy.

Using \eqref{eq:relativeentropy} and \eqref{eq:MUI}, the waveform design problem can be formulated as
\begin{align} \label{eq:WaveformDesign}
\max\limits_{\bX} & \ \log\det(\bR_1) + \tr[\bR_1^{-1}(\widetilde{\bX}\bg_d\bg_d^\HH\widetilde{\bX}^\HH+\sigma^2\bI_{N_\textrm{RL}})]\nonumber\\
\textrm{s.t.} &\ \|\bH\bX^\T-\bS\|^2_\textrm{F}\leq \epsilon, \nonumber\\
&\ \textrm{tr}(\bX\bX^\HH)\leq P_\textrm{t},
\end{align}
where the constant terms in the objective function have been ignored.

\section{Waveform Design Based on MM} \label{Sec:AlgorithmDesignMM}
To tackle the optimization problem in \eqref{eq:WaveformDesign}, we leverage the MM method and minorize the highly nonlinear objective function with a simpler function. The minorized function (also called the minorizer or the surrogate function) of the objective at the $(k+1)$-th iteration, denoted by $ Q(\bX;\bX_k)$, should satisfy
\begin{subequations}
  \begin{align}
  Q(\bX;\bX_k) & \leq g(\bX), \\
  Q(\bX_k;\bX_k) & = g(\bX_k),
\end{align}
\end{subequations}
where $g(\bX)$ is the objective function in \eqref{eq:WaveformDesign}, and $\bX_k$ is the solution at the $k$th iteration.

For simplicity and without loss of generality, we assume that the noise power is $\sigma^2=1$. Let $\bR_{X}=\bX^\HH\bX$ be the waveform covariance matrix, and let $\widetilde{\bR}_{X}=\widetilde{\bX}^\HH\widetilde{\bX}=\bI_{N_\textrm{R}}\otimes \bR_{X}$.
Using the identity $\det(\bI+\bA\bB)=\det(\bI+\bB\bA)$ \cite{HornMatrix1990}, we rewrite $\log\det(\bR_1)$ as
\begin{align} \label{eq:block}
\log\det(\bR_1)=&\log\det(\bR_\textrm{G}^{\frac{1}{2}}\widetilde{\bR}_{X}\bR_\textrm{G}^{\frac{1}{2}}+\bI) \nonumber \\
=&\log\det(\bE\bC^{-1}\bE^\HH),
\end{align}
where $\bE=[\bI_{N_{\textrm{TR}}},\textbf0_{N_{\textrm{TR}}\times N_{\textrm{RL}}}]$,
$$\bC=
\begin{bmatrix}
\bI_{N_{\textrm{TR}}}&\bR_\textrm{G}^{\frac{1}{2}}\widetilde{\bX}^\HH \\
\widetilde{\bX}\bR_\textrm{G}^{\frac{1}{2}} & \bR_1
\end{bmatrix},$$
and we have employed the block matrix inversion lemma \cite{HornMatrix1990} in the second line of \eqref{eq:block}.

$\log\det(\bE\bC^{-1}\bE^\HH)$ is convex with respect to $\bC$ \cite{TangMinorization2018}. According to the property of convex functions \cite{BoydConvex2004}, we obtain that
\begin{align} \label{eq:PartI}
\log\det(\bE\bC^{-1}\bE^\HH)\geq
&\log\det(\bE\bC_{k}^{-1}\bE^\HH)
\nonumber\\&
+\textrm{tr}[\bT_k(\bC-\bC_{k})],
\end{align}
where $\bT_k=-\bC_{k}^{-1}\bE^\HH(\bE\bC_{k}^{-1}\bE^\HH)^{-1}\bE\bC_{k}^{-1}$ is the gradient of $\log\det(\bE\bC^{-1}\bE^\HH)$ at $\bC_{k}$ \cite{HjorungnesDifferentiation2007}. Partition $\bT_k$ as
\begin{equation}
\bT_k=
\begin{bmatrix}
\bT_{k}^{11} & \bT_{k}^{12} \\
(\bT_{k}^{12})^\HH & \bT_{k}^{22}
\end{bmatrix},
\end{equation}
where $\bT_{k}^{11}\in\mathbb{C}^{N_{\textrm{TR}}\times N_{\textrm{TR}}}$, $\bT_{k}^{12}\in\mathbb{C}^{N_{\textrm{TR}}\times N_{\textrm{RL}}}$, and $\bT_{k}^{22}\in\mathbb{C}^{N_{\textrm{RL}}\times N_{\textrm{RL}}}$. We have
\begin{align} \label{eq:Tl}
\textrm{tr}(\bT_k\bC)=
&c_{k}+2\Re[\textrm{tr}(\widetilde{\bX}\bR_\textrm{{G}}^{\frac{1}{2}}\bT_{k}^{12})]
\nonumber\\
&+\textrm{tr}(\bT_{k}^{22}\widetilde{\bX}\bR_\textrm{{G}}\widetilde{\bX}^\HH),
\end{align}
where $c_{k}=\textrm{tr}(\bT_{k}^{11}+\bT_{k}^{22})$ is a constant, which does not depend on $\bX$.

According to \cite{Wang2022Rician}, a minorizer of the objective in \eqref{eq:WaveformDesign} can be given by
\begin{align} \label{eq:kthIteration}
&2\Re[\textrm{tr}(\widetilde{\bX}^\HH \bP_k)]
+\textrm{tr}(\bQ_k\widetilde{\bX}\bR_\textrm{G}\widetilde{\bX}^\HH)+c_1,
\end{align}
where
\begin{align}
\bP_k = (\bT_k^{12})^\HH\bR_\textrm{G}^{\frac{1}{2}}+\bR_{1,k}^{-1}(\widetilde{\bX}_k\bg_d)\bg_d^\HH,
\end{align}
\begin{align}
\bQ_k = \bT^{22}_k-\bR_{1,k}^{-1}(\widetilde{\bX}_k\bg_d)(\widetilde{\bX}_k\bg_d)^\HH\bR_{1,k}^{-1}-\bR_{1,k}^{-2},
\end{align}
$\bR_{1,k} = \widetilde{\bX}_k\bR_\textrm{G}\widetilde{\bX}_k^\HH+\bI$, and $c_1$ is a constant term.
The expression in \eqref{eq:kthIteration} can be rewritten as
\begin{equation} \label{eq:objective}
\widetilde{\bx}^\HH\widetilde{\bM}_k\widetilde{\bx}+2\textrm{Re}(\widetilde{\bx}^\HH\widetilde{\bm}_k)+c_1,
\end{equation}
where $\widetilde{\bx}=\textrm{vec}(\widetilde{\bX}),\widetilde{\bM}_k=\bR_\textrm{G}^*\otimes\bQ_k,\widetilde{\bm}_k=\textrm{vec}(\bP_k)$. Note that  $\widetilde{\bx} = \bG_s\bx$ \cite{TangMinorization2018}, where $\bx=\textrm{vec}(\bX)$, $\bG_s=\bJ_s\otimes\bI_{L}$, $\bJ_s=[\bJ_1,\bJ_2,\cdots,\bJ_{N_{\textrm{TR}}}]^\T$, with $\bJ_i,i=1,2,\cdots,N_{\textrm{TR}}$, representing an $N_\textrm{T}\times N_\textrm{R}$ elementary matrix, whose $(i_r,i_c)${th} element is 1, and all other elements are 0, $i_r=1+\textrm{mod}(i-1,N_\textrm{T})$, and $i_c=\lceil \frac{i}{N_\textrm{T}}\rceil$.

The MUI energy constraint in \eqref{eq:WaveformDesign} is equivalent to $\|\bX\bH^\T-\bS^\T\|^2_\textrm{F} \leq \epsilon$. Using the identity that $\textrm{vec}(\bA\bB\bC)=(\bC^\T\otimes\bA)\textrm{vec}(\bB)$, we can obtain that $\textrm{vec}(\bX\bH^\T)=\widetilde{\bH}\bx$, where $\widetilde{\bH}=\bH\otimes\bI_L$. Then the MUI energy constraint can be rewritten as
\begin{equation}
\|\widetilde{\bH}\bx-\bs\|^2_2\leq \epsilon,
\end{equation}
where $\bs=\textrm{vec}(\bS^\T)$.

Then at the $(k+1)$-th iteration of the MM method, the minorized problem based on \eqref{eq:objective} can be expressed as
\begin{align} \label{eq:QCQP}
\max_\bx \ &\bx^\HH\bM_k\bx+2\textrm{Re}(\bx^\HH\bm_k) \nonumber\\
\textrm{s.t.}\ & \|\widetilde{\bH}\bx-\bs\|^2_2\leq \epsilon, \nonumber\\
& \bx^\HH\bx\leq P_\textrm{t},
\end{align}
where $\bM_k=\bG_s^\HH\widetilde{\bM}_k\bG_s$, $\bm_k=\bG_s^\HH\widetilde{\bm}_k$, and we have used the fact that $\textrm{tr}(\bX\bX^\HH)=\bx^\HH\bx$.

Next, we present two methods to tackle the QCQP problem in \eqref{eq:QCQP}.
\subsection{SDR followed by a rank-one decomposition procedure} \label{Sec:MM-SDR}

First, we introduce an auxiliary variable $t \in \mathbb{C}$, which satisfies $|t|^2=1$.
Let $\widehat{\bx}= [\bx^\top  \ t]^\top $,
$$
\widehat{\bH}=
\begin{bmatrix}
\widetilde{\bH}^\HH\widetilde{\bH} &-\widetilde{\bH}^\HH\bs\\
-\bs^\HH\widetilde{\bH} &\bs^\HH\bs
\end{bmatrix}.$$
Then the MUI energy constraint can be rewritten as
$$\widehat{\bx}^\HH\widehat{\bH}\widehat{\bx}\leq \epsilon .$$
Define $\bJ_1=\textrm{blkDiag}({\bI}_{ N_\textrm{T}L};0)$, $\bJ_0 = \textrm{blkDiag}(\bzero_{ N_\textrm{T}L};1)$, and $$\widehat{\bM}_k = \begin{bmatrix}
{\bM}_k &{\bm}_k\\
{\bm}_k^\HH &0
\end{bmatrix}.$$

Then the optimization problem in \eqref{eq:QCQP} can be reformulated as
\begin{align} \label{eq:Optquadratic22}
\max_{\widehat{\bx}} \ &\widehat{\bx}^\HH\widehat{\bM}_k\widehat{\bx} \nonumber\\
\textrm{s.t.}\ & \|\bJ_1\widehat{\bx}\|^2_2 \leq P_\textrm{t}, \|\bJ_0\widehat{\bx}\|^2_2 = 1,\nonumber\\
& \widehat{\bx}^\HH\widehat{\bH}\widehat{\bx}\leq \epsilon.
\end{align}

It is worth noting that the problem in \eqref{eq:Optquadratic22} is a hidden convex optimization problem \cite{Huang2007Complex} and can be tackled by the SDR method \cite{Luo2010Semidefinite}. Specifically, let $\widehat{\bX} = \widehat{\bx}\widehat{\bx}^\HH$, and we recast the optimization problem in \eqref{eq:Optquadratic22} as the following optimization problem:
\begin{align} \label{eq:Optquadratic23}
\max_{\widehat{\bX}} \ &\textrm{tr}(\widehat{\bM}_k\widehat{\bX}) \nonumber\\
\textrm{s.t.}\ & \textrm{tr}(\bJ_1\widehat{\bX}) \leq P_\textrm{t}, \textrm{tr}(\bJ_0\widehat{\bX}) = 1, \nonumber\\
& \textrm{tr}(\widehat{\bH}\widehat{\bX})\leq \epsilon, \textrm{rank}(\widehat{\bX}) = 1, \widehat{\bX}\succeq \mathbf{0}.
\end{align}
Dropping the rank constraint in \eqref{eq:Optquadratic23}, we obtain the following relaxed optimization problem:

\begin{align} \label{eq:Optquadratic4}
\max_{\widehat{\bX}} \ &\textrm{tr}(\widehat{\bM}_k\widehat{\bX}) \nonumber\\
\textrm{s.t.}\ & \textrm{tr}(\bJ_1\widehat{\bX}) \leq P_\textrm{t},
 \textrm{tr}(\bJ_0\widehat{\bX}) = 1, \nonumber\\
 & \textrm{tr}(\widehat{\bH}\widehat{\bX})\leq \epsilon,  \widehat{\bX}\succeq \mathbf{0}.
\end{align}

It can be checked that the optimization problem in \eqref{eq:Optquadratic4} is convex (more precisely, a semi-definite programming (SDP) problem). Thus, it can be solved by the interior point method (e.g.,  via the CVX toolbox \cite{Stephen2008CVX}). Let $\widehat{\bX}^\star$ represent the solution to problem \eqref{eq:Optquadratic4}. Then the optimal $\widehat{\bx}^\star$ can be obtained by a rank-one decomposition procedure. Specifically, according to \cite[Theorem 2.3]{Ai2011Rank_one}, we can find a nonzero vector $\widehat{\bx}^\star$, which satisfies $({\widehat{\bx}^\star})^\HH\widehat{\bM}_k\widehat{{\bx}}^\star=\textrm{tr}(\widehat{\bM}_k\widehat{\bX}^\star)$, $({\widehat{\bx}^\star})^\HH{\bJ}_1\widehat{{\bx}}^\star=\textrm{tr}({\bJ}_1\widehat{\bX}^\star)$, $({\widehat{\bx}^\star})^\HH{\bJ}_0\widehat{{\bx}}^\star=\textrm{tr}({\bJ}_0\widehat{\bX}^\star)$ and $({\widehat{\bx}^\star})^\HH\widehat{\bH}\widehat{{\bx}}^\star=\textrm{tr}(\widehat{\bH}\widehat{\bX}^\star)$. We can check that $\widehat{{\bx}}^\star({\widehat{\bx}^\star})^\HH$ is optimal for both \eqref{eq:Optquadratic23} and \eqref{eq:Optquadratic4}. Therefore, ${\widehat{\bx}^\star}$ is the optimal solution to \eqref{eq:Optquadratic22}.
Let $\bx^\star=\overline{\bx}^\star/u$, where $\overline{\bx}^\star$ is the vector containing the first $N_\textrm{T}L$ elements of $\widehat{\bx}^\star$, and $u$ is the last element of $\widehat{\bx}^\star$. It is easy to verify that $\bx^\star$ not only maximizes the objective in \eqref{eq:QCQP}, but also satisfies the constraints. Therefore, $\bx^\star$ is the optimal solution to \eqref{eq:QCQP}.
\subsection{ADMM} \label{Sec:MM-ADMM}

Note that the computational complexity of solving an SDP problem is high. To reduce the computational burden, we derive an ADMM algorithm to tackle the optimization problem \eqref{eq:QCQP}.
To proceed, we introduce an auxiliary variable $\bz$, and recast the optimization problem in \eqref{eq:QCQP} as
\begin{align} \label{eq:Optquadratic2}
\min_{{\bx},\bz} \ &\bx^\HH\overline{\bM}_k\bx+2\textrm{Re}(\bx^\HH\overline{\bm}_k) \nonumber\\
\textrm{s.t.}\ & {\bz}=\widetilde{\bH}\bx-\bs, \|\bz\|^2_2\leq \epsilon,\ \nonumber\\
&  \bx^\HH\bx\leq P_\textrm{t},
\end{align}
where $\overline{\bM}_k=-\bM_k$, and $\overline{\bm}_k=-\bm_k$.
The augmented Lagrangian associated with \eqref{eq:Optquadratic2} is given by
\begin{align}
L_\mu({\bx},\bz,\blambda)=&{\bx}^\HH\overline{\bM}{\bx}+2\textrm{Re}({\bx}^\HH\overline{\bm}) \nonumber\\
&+\frac{\mu}{2}\{\|\bz-\widetilde{\bH}{\bx}+\bs+\mathbf{\blambda}\|^2_2-\|{\blambda}\|^2_2\},
\end{align}
where $\mu$ is the penalty parameter, $\blambda$ is the Lagrangian multiplier, and to lighten the notations, we omit the subscript $k$.

At the $(m+1)$-th iteration of the ADMM algorithm, we carry out the following steps:
\begin{align}
{\bx}_{m+1}=&\arg \min\limits_{{\bx}}L_\mu({\bx},\bz_m,\blambda_m), \label{eq:xm} \\
\bz_{m+1} =&\arg\min\limits_{\bz}L_\mu({\bx}_{m+1},\bz,\blambda_m), \label{eq:zmtm} \\
\blambda_{m+1}=&\blambda_{m}+\bz_{m+1}-\widetilde{\bH}{\bx}_{m+1}+\bs. \label{eq:d_m}
\end{align}

$\bullet$ \textbf{Solution to \eqref{eq:xm}:}

The optimization problem in \eqref{eq:xm} can be written as
\begin{align} \label{eq:auxiliary}
\min\limits_{{\bx}}\ &{\bx}^\HH{\bB}{\bx}+2\textrm{Re}({\bx}^\HH\bb_m)\nonumber\\
\textrm{s.t.}\ & \bx^\HH\bx\leq P_\textrm{t},
\end{align}
where $\bB=\overline{\bM}+\frac{\mu}{2}\widetilde{\bH}^\HH\widetilde{\bH}$, and $\bb_m=\overline{\bm}-\frac{\mu}{2}\widetilde{\bH}^\HH(\bz_m+\bs+\blambda_m)$.
The optimization problem in \eqref{eq:auxiliary} is hidden-convex and can be solved by the Lagrange multiplier method. Let the Lagrangian associated with \eqref{eq:auxiliary} be
\begin{equation}
F(\bx,\nu)={\bx}^\HH{\bB}{\bx}+2\textrm{Re}({\bx}^\HH\bb_m)+\nu({\bx}^\HH{\bx}-P_\textrm{t}),
\end{equation}
where $\nu$ is the Lagrange multiplier associated with the constraint in \eqref{eq:auxiliary}. By differentiating \eqref{eq:auxiliary} w.r.t. $\bx$ and setting the differentiation equal to zero, the minimizer is given by
\begin{equation} \label{eq:Updatexm}
\bx_{m+1}=-(\bB+\nu_{m+1}\bI_{N_\textrm{T}L})^{-1}\bb_m,
\end{equation}
where $\nu_{m+1}$ is the solution to the following equation:
\begin{equation}
\bb_{m}^\HH(\bB+\nu_{m+1}\bI_{N_\textrm{T}L})^{-2}\bb_m=P_\textrm{t}.
\end{equation}
$\bullet$ \textbf{Solution to \eqref{eq:zmtm}:}

The optimization problem in \eqref{eq:zmtm} can be written as
\begin{align} \label{eq:z_dm}
\min_\bz\ & \|\bz-\bp_m\|^2_2 \nonumber\\
\textrm{s.t.}\ & \|\bz\|^2_2\leq \epsilon,
\end{align}
where $\bp_m={\widetilde{\bH}\x}_{m+1}-\bs-\blambda_m$. The optimal solution to \eqref{eq:z_dm} is given by
\begin{equation} \label{eq:z_Update}
\bz_{m+1}=\left\{
\begin{aligned}
&\bp_m,& \textrm{if}\ \|\bp_m\|^2_2\leq \epsilon, \\
&\sqrt{\epsilon}\bp_m/\|\bp_m\|_2,& \textrm{if}\ \|\bp_m\|^2_2> \epsilon.
\end{aligned}
\right.
\end{equation}

We summarize the presented ADMM algorithm (for the problem in \eqref{eq:QCQP}) in Algorithm \ref{Alg:1}, in which the ADMM algorithm is terminated when  $\|\br_m\|_2\leq \varepsilon^{\textrm{primal}}_1$ and $\|\bd_m\|_2\leq \varepsilon^{\textrm{dual}}_1$, where $\br_m=\widetilde{\bH}{\bx}_{m}-\bs-\bz_{m}$, $\bd_m=\bx_m-\bx_{m-1}$ , $\varepsilon^{\textrm{primal}}_1$ and $\varepsilon^{\textrm{dual}}_1$ are user-defined small values.
\begin{algorithm}[!htbp]
 \caption{ \small  ADMM Algorithm for the QCQP Problem in \eqref{eq:QCQP}.}\label{Alg:1}
  \KwIn{$\bM_k$, $\bm_k$, $\bH$, $\bs$, $P_\textrm{t}$, $\epsilon$.}
  \KwOut{$\bx_{k+1}$.}
  \textbf{Initialize:} $m=0,\bx_{k,0}=\bx_k,\bz_{k,0}=\bx_{k,0},\blambda_{k,0}=\textbf{0}$.\\
  $\bB_k=\overline{\bM}_k+\frac{\mu}{2}\widetilde{\bH}^\HH\widetilde{\bH}$.\\
   \Repeat{convergence}{$\bb_{k,m}=\overline{\bm}_{k}-\frac{\mu}{2}\widetilde{\bH}^\HH\bz_{k,m}+\bs+\blambda_{k,m}$; \\
   $\bx_{k,m+1}=-(\bB_k+\nu_{k,m+1}\bI_{N_\textrm{T}L})^{-1}\bb_{k,m}$; \\
   $\bp_{k,m}={\widetilde{\bH}\x}_{k,m+1}-\bs-\blambda_{k,m}$;\\
   $\bz_{k,m+1}=\textrm{min}(\sqrt{\epsilon}/\|\bp_{k,m}\|_2,1)\cdot\bp_{k,m}$; \\
   $\blambda_{k,m+1}=\blambda_{k,m}+\bz_{k,m+1}-\widetilde{\bH}{\bx}_{k,m+1}+\bs$; \\
   $m=m+1$;\\
   $\br_{k,m}=\widetilde{\bH}{\bx}_{k,m}-\bs-\bz_{k,m}$; \\
   $\bd_{k,m}=\bx_{k,m}-\bx_{k,m-1}$;
   }
   $\bx_{k+1}=\bx_{k,m}$.
\end{algorithm}

\begin{algorithm}[!htbp]
 \caption{ \small  MM-based Waveform Design for Rician Target Detection with MIMO DFRC Systems.}\label{Alg:2}
  \KwIn{$\bg_d$, $\bR_\textrm{G}$, $\bH$, $\bs$, $P_\textrm{t}$, $\epsilon$.}
  \KwOut{$\bx^{\star}$.}
  \textbf{Initialize:} $k=0,\bX_{0}$  (e.g., with quasi-orthogonal waveforms).\\
  \Repeat{convergence}{
  $\bR_{1,k} = \widetilde{\bX}_k\bR_\textrm{G}\widetilde{\bX}_k^\HH+\bI$; \\
    $\bQ_k = \bT^{22}_k-\bR_{1,k}^{-1}(\widetilde{\bX}_k\bg_d)(\widetilde{\bX}_k\bg_d)^\HH\bR_{1,k}^{-1}-\bR_{1,k}^{-2};$\\
    $\bP_k = (\bT_k^{12})^\HH\bR_\textrm{G}^{\frac{1}{2}}+\bR_{1,k}^{-1}(\widetilde{\bX}_k\bg_d)\bg_d^\HH;$\\
    $\widetilde{\bM}_k=\bR_\textrm{G}^*\otimes\bQ_k$ ;\\
    $\widetilde{\bm}_k=\textrm{vec}(\bP_k)$;\\
   $\bM_k=\bG_s^\HH\widetilde{\bM}_k\bG_s$;\\
   $\bm_k=\bG_s^\HH\widetilde{\bm}_k$;\\
   Update $\bx_{k+1}$ by the SDR approach or the ADMM algorithm in Algorithm \ref{Alg:1}; \\
   $k=k+1$.\\
  }
  $\bx^{\star}=\bx_{k}$.
\end{algorithm}
\subsection{Analysis and Algorithm Extensions}
The  waveform design algorithm based on MM is summarized in Algorithm \ref{Alg:2}. The presented algorithm is terminated if $|D_k-D_{k-1}|/D_k<\xi_1$, where $D_k$ denotes the relative entropy at the $k$-th iteration, and $\xi_1$ is a small user-defined value. Next we provide the per-iteration computational complexity of the presented algorithm. Table II presents the details of the computation load for the presented algorithm.  We can observe that for typical values of $N_\textrm{T}$, $M$ and $L$, the computational complexity of solving the QCQP problem in  \eqref{eq:QCQP} based on ADMM is lower than that based on SDR. 

\begin{table}[!htbp] \label{tab:2}
\caption{Computational complexity analysis.}
\begin{center}
\begin{tabular}{cc|cc}
\hline
\multicolumn{2}{c|}{Computation}  & \multicolumn{2}{c}{Complexity}\\
\hline
\multicolumn{2}{c|}{$\bM_k$}  & \multicolumn{2}{c}{$O(N_\textrm{RL}^3+(N_\textrm{TR}N_\textrm{RL})^2)$}\\
\hline
\multicolumn{2}{c|}{$\bm_k$}  & \multicolumn{2}{c}{$O(N_\textrm{RL}^3+N_\textrm{RL}^2 N_\textrm{TR})$}\\
\hline
\multicolumn{4}{c}{\emph{\textsf{Solving the QCQP problem in \eqref{eq:QCQP}}}}\\
\hline
\multicolumn{2}{c|}{SDR}  & \multicolumn{2}{c}{ADMM (each inner loop)}\\
\hline
Computation & Complexity & Computation & Complexity \\
\hline
Solving \eqref{eq:Optquadratic4} & $O((N_\textrm{T}L)^{4.5})$ & $\bx_m$ & $O((N_\textrm{T}L)^3+MN_\textrm{T}L^2)$ \\
\hline
\multirow{2}*{\makecell[b]{Rank-one\\ decomposition}}& \multirow{2}*{$O((N_\textrm{T}L)^{3})$} & $\bz_m$ & $O(MN_\textrm{T}L^2)$ \\
\cline{3-4}
~&~ & $\blambda_m$ & $O(MN_\textrm{T}L^2)$ \\
\hline
Total & $O((N_\textrm{T}L)^{4.5})$ & Total & $O((N_\textrm{T}L)^3+MN_\textrm{T}L^2)$ \\
\hline
\end{tabular}
\end{center}
\end{table}

We also note that the proposed ADMM algorithm not only have a lower computational complexity, but also can be extended to deal with other constraints. In practice, to avoid the waveform distortion caused by nonlinear amplifiers, we will enforce a PAPR constraint on the waveforms:
\begin{align*}
\bx_n^\HH\bx_n=P_\textrm{t}/N_\textrm{T}, \textrm{PAPR}(\bx_n)\leq\rho, n=1,2\ldots,N_\textrm{T},
\end{align*}
where $1\leq\rho\leq L$,
\begin{align*}
\textrm{PAPR}(\bx_n)=\frac{\max_l|x_n(l)|^2}{\frac{1}{L}\sum_{l=1}^L|x_n(l)|^2}, n=1,2\ldots,N_\textrm{T},
\end{align*}
and $x_n(l)$ is the $l$-th element of $\bx_n$, $l=1,2,\cdots,L$. To extend the presented ADMM algorithm to deal with the PAPR constraint, we only need to replace the optimization problem in \eqref{eq:auxiliary} by
\begin{align} \label{eq:auxiliaryCM}
\min\limits_{{\bx}}\ &{\bx}^\HH{\bB}{\bx}+2\textrm{Re}({\bx}^\HH\bb_m)\nonumber\\
\textrm{s.t.}\ & \bx_n^\HH\bx_n=P_\textrm{t}/N_\textrm{T}, \textrm{PAPR}(\bx_n)\leq\rho, n=1,2\ldots,N_\textrm{T}.
\end{align}
This optimization problem can be tackled efficiently by the MM method proposed in \cite{Tang2021RangeSpread}.

\section{Waveform Design Based on ADMM} \label{Sec:AlgorithmDesignADMM}
Next, we present a novel ADMM algorithm to tackle the non-convex problem in \eqref{eq:WaveformDesign}. To this purpose, we use the variable splitting trick, and rewrite the optimization problem in \eqref{eq:WaveformDesign} as
\begin{align} \label{eq:splitting_U}
\max\limits_{\bX,\bU} & \ \log\det(\bR_1) + \tr[\bR_1^{-1}(\widetilde{\bX}\bg_d\bg_d^\HH\widetilde{\bX}^\HH+\bI_{N_\textrm{RL}})]\nonumber\\
\textrm{s.t.} &\ \bU=\bX\bH^\T-\bS^\T,  \|\bU\|^2_\textrm{F} \leq \epsilon, \nonumber\\
&\ \textrm{tr}(\bX\bX^\HH)\leq P_\textrm{t},
\end{align}
where $\bU$ is the introduced auxiliary variable. The augmented Lagrangian  associated with \eqref{eq:splitting_U} is given by
\begin{align} \label{eq:ADMM}
L_\nu({\bX},\bU,\bLam)=& -\log\det(\bR_1) - \tr[\bR_1^{-1}(\widetilde{\bX}\bg_d\bg_d^\HH\widetilde{\bX}^\HH+\bI_{N_\textrm{RL}})]\nonumber\\
&+\frac{\nu}{2}[\|\bU-\bX\bH^\T+\bS^\T+\bLam\|^2_\textrm{F}-\|\bLam\|^2_\textrm{F}],
\end{align}
where $\nu$ is the penalty parameter.

At the $(m+1)$-th iteration of the ADMM algorithm, we carry out the following steps:
\begin{align}
{\bX}_{m+1}=&\arg \min\limits_{{\bX}}L_\nu({\bX},\bU_m,\bLam_m), \label{eq:Xm} \\
\bU_{m+1} =&\arg\min\limits_{\bU}L_\nu({\bX}_{m+1},\bU,\bLam_m), \label{eq:Um} \\
\bLam_{m+1}=&\bLam_{m}+\bU_{m+1}-{\bX}_{m+1}{\bH}^\T+\bS^\T. \label{eq:Lambdam}
\end{align}

$\bullet$ \textbf{Solution to \eqref{eq:Xm}:}

The optimization problem in \eqref{eq:Xm} is formulated as (For notation simplicity, we omit the subscript $m$ in the following derivations.)
\begin{align} \label{eq:2Part}
\max\limits_{\bX} & \ \underbrace{\log\det(\bR_1)+ \tr[\bR_1^{-1}(\widetilde{\bX}\bg_d\bg_d^\HH\widetilde{\bX}^\HH+\bI_{N_\textrm{RL}})]}_{\textrm{Part I}}\nonumber\\
&-\underbrace{\frac{\mu}{2}\|\bU-\bX\bG^\T+\bS^\T+\bLam\|^2_\textrm{F}}_{\textrm{Part II}}\nonumber\\
\textrm{s.t.} &\ \textrm{tr}(\bX\bX^\HH)\leq P_\textrm{t}.
\end{align}
Again, we leverage the MM method to tackle the problem in \eqref{eq:2Part}. In Section \ref{Sec:AlgorithmDesignMM}, we have derived that a minorizer of Part I in \eqref{eq:2Part} is given by $2\Re[\textrm{tr}(\widetilde{\bX}^\HH \bP_k)]
+\textrm{tr}(\bQ_k\widetilde{\bX}\bR_\textrm{G}\widetilde{\bX}^\HH)+c_1$. Thus, the minorized problem of \eqref{eq:2Part} at the $(k+1)$-th (inner) iteration can be formulated as
\begin{align} \label{eq:MM2Part}
\max\limits_{\bX} & \ \textrm{tr}(\bQ_k\widetilde{\bX}\bR_\textrm{G}\widetilde{\bX}^\HH)
+2\Re[\textrm{tr}(\widetilde{\bX}^\HH \bP_k)] \nonumber\\
&-\frac{\mu}{2}\|\bU-\bX\bH^\T+\bS^\T+\bLam\|^2_\textrm{F}\nonumber\\
\textrm{s.t.} &\ \textrm{tr}(\bX\bX^\HH)\leq P_\textrm{t}.
\end{align}

Note that the minorized problem in \eqref{eq:MM2Part} can be rewritten as
\begin{align} \label{eq:auxiliary2}
\max\limits_{{\bx}}\ &{\bx}^\HH{\bA_k}{\bx}+2\textrm{Re}({\bx}^\HH\ba_k)\nonumber\\
\textrm{s.t.}\ & \bx^\HH\bx \leq P_\textrm{t},
\end{align}
where $\bA_k={\bM}_k-\frac{\mu}{2}\widetilde{\bH}^\HH\widetilde{\bH}, \ba_k={\bm}_k+\frac{\mu}{2}\widetilde{\bH}^\HH(\bu+\bs+\blambda)$, $\bu=\vec(\bU)$, and $\blambda=\vec(\bLam)$. Similarly, the optimization problem in \eqref{eq:auxiliary2} can be tackled by the Lagrange multiplier method.

$\bullet$ \textbf{Solution to \eqref{eq:Um}:}

The optimization problem in \eqref{eq:Um} can be written as
\begin{align} \label{eq:UD}
\min_\bU\ & \|\bU-\bV\|^2_\textrm{F} \nonumber\\
\textrm{s.t.}\ & \|\bU\|^2_\textrm{F}\leq \epsilon,
\end{align}
where $\bV=\bX\bH^\T-\bS^\T-\bLam$. The optimal solution to \eqref{eq:UD} is given by
\begin{equation}
\bU=\left\{
\begin{aligned}
&\bV,& \textrm{if}\ \|\bV\|^2_\textrm{F}\leq \epsilon, \\
&\sqrt{\epsilon}\bV/\|\bV\|_\textrm{F},& \textrm{if}\ \|\bV\|^2_\textrm{F}> \epsilon.
\end{aligned}
\right.
\end{equation}

We summarize the proposed ADMM algorithm in Algorithm \ref{Alg:3}. The ADMM algorithm is terminated when  $\|\bR_m\|_\textrm{F}\leq \varepsilon^{\textrm{primal}}_2$ and $\|\bD_m\|_\textrm{F}\leq \varepsilon^{\textrm{dual}}_2$, where $\bR_m={\bX}_{m}{\bH}^\T-\bU_{m}-\bS^\T$, $\bD_m=\bX_m-\bX_{m-1}$ , $\varepsilon^{\textrm{primal}}_2$ and $\varepsilon^{\textrm{dual}}_2$ are user-defined small values. Moreover, we highlight that the ADMM algorithm can be easily extended to design PAPR-constrained waveforms (i.e., replace the energy constraint in \eqref{eq:auxiliary2} by the PAPR constraint and tackle the associated waveform design problem by the MM method in \cite{Tang2021RangeSpread}).

\begin{algorithm}[!htp]
 \caption{ \small  ADMM-based Waveform Design for Rician Target Detection with MIMO DFRC Systems.}\label{Alg:3}
  \KwIn{$\bg_d$, $\bR_\textrm{G}$, $\bH$, $\bS$, $P_\textrm{t}$, $\epsilon$}
  \KwOut{$\bX^{\star}$}
  \textbf{Initialize:} $m=0,\bX_{0}$ (e.g. with quasi-orthogonal waveforms),$\bU_{0}=\bX_{0},\bLam_{0}=\bzero.$\\
  \Repeat{convergence}{
   Update $\bX_{m+1}$ with MM method; \\
   $\bV_m=\bX_{m+1}\bH^\T-\bS^\T-\bLam_{m}$;\\
   $\bU_{m+1}=\textrm{min}(\sqrt{\epsilon}/\|\bV_{m}\|_\textrm{F},1)\cdot\bV_{m}$; \\
   $\bLam_{m+1}=\bLam_{m}+\bU_{m+1}-{\bX}_{m+1}{\bH}^\T+\bS^\T$; \\
   $m=m+1$;\\
   $\bR_m={\bX}_{m}{\bH}^\T-\bU_{m}-\bS^\T$; \\
   $\bD_m=\bX_m-\bX_{m-1}$;
   \\}
   $\bX^{\star}=\bX_{m}$.
\end{algorithm}

\section{Numerical Examples} \label{Sec:NumericalExamples}
In this section, we provide examples to verify the performance of the presented algorithms. Consider a MIMO DFRC system with $N_\textrm{T}=12$ transmitters and $N_\textrm{R}=12$ receivers. The inter-element spacings of the antennas is $\lambda/2$ ($\lambda$ is the wavelength). The code length is $L=10$ and the transmit energy is $P_\textrm{t} = 1$. The Rician target is composed of a deterministic scatterer and 30 random scatterers. The DOA of the deterministic scatterer is $\theta_d=15^{\circ}$ and $\bg_d$ is modeled as $\bg_d=\alpha_d\ba(\theta_d)\otimes \bb(\theta_d)$ with $\alpha_d=\sqrt{3/2}$. The DOAs of the random scatterers are uniformly distributed from $-60^{\circ}$ to $56^{\circ}$, and the amplitudes of the random scatterers are independent random variables, obeying a circularly-symmetric Gaussian distribution with zero mean and variance $\sigma^2_r=0.05$. In other words, $\bR_\textrm{G}$ is modeled by
\begin{equation}
\bR_\textrm{G}=\sum_k\sigma^2_r(\bb(\theta_k)\bb^\HH(\theta_k))\otimes(\ba(\theta_k)\ba^\HH(\theta_k)).
\end{equation}
$M=4$ communication users are to be served, and the maximum allowed MUI energy of the communication signals is $10^{-6}$. The desired communication signals for these users are quadrature phase shift keying (QPSK) signals with the energy of $e_\textrm{c}=0.1$. The entries of the channel matrix $\bH$ are i.i.d. Gaussian random variables, with zero mean and variance of $1$ (i.e., we consider a flat fading channel). The noise power is $\sigma^2=1$. The proposed algorithms are initialized with quasi-orthogonal waveforms. For the presented algorithms, we set $\xi_1=\xi_2=10^{-4}$ when using MM methods. For the ADMM algorithm, we set $\varepsilon^{\textrm{primal}}_k=10^{-6}$ and $\varepsilon^{\textrm{dual}}_k=10^{-3}$ ($k=1,2$), respectively. 

\begin{figure}[!htbp]
\centering
{{{\includegraphics[width = 0.4\textwidth]{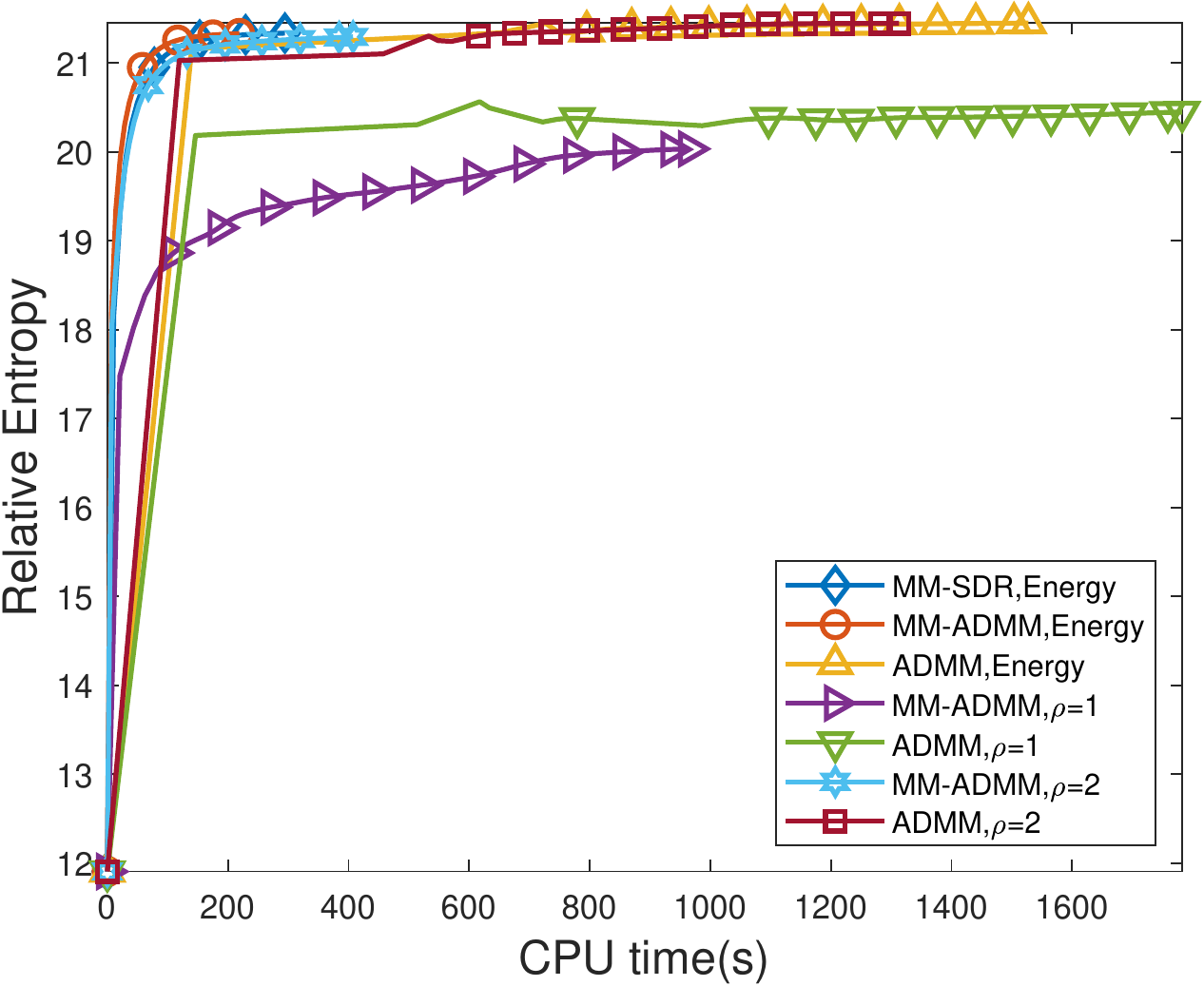}}} }
\caption{Relative entropy of the synthesized waveforms versus the CPU time ($P_\textrm{t} = 1$). }
\label{Fig:2}
\end{figure}

First, we analyze the convergence of the presented algorithms.  Fig. \ref{Fig:2} shows the relative entropy of the waveforms synthesized under the energy constraint and the PAPR constraint by the presented algorithms versus the CPU time. Note that when we use the proposed MM algorithm to synthesize the energy-constrained waveforms, the SDR approach in Section \ref{Sec:MM-SDR} and the ADMM approach in Section \ref{Sec:MM-ADMM} can be utilized to tackle the QCQP problem in \eqref{eq:QCQP}, respectively. Thus, we name the associated algorithms MM-SDR and MM-ADMM, respectively.
The results show that all the presented algorithms (MM-SDR, MM-ADMM, and ADMM) converge to a finite value after a number of iterations \footnote{We note that MM-SDR cannot be extended to deal with a PAPR constraint.}.
In addition, the waveforms synthesized by the MM-ADMM algorithm can achieve the same performance as those synthesized by the MM-SDR algorithm. This implies that if only the energy constraint is imposed, the ADMM approach in Section \ref{Sec:MM-ADMM} can obtain a globally optimal solution to the QCQP problem in \eqref{eq:QCQP}. Note that the MM-ADMM algorithm converges in a shorter time than the MM-SDR algorithm.  Therefore, in the following analysis, we mainly use the ADMM approach to tackle the QCQP problem encountered at each iteration. Fig. \ref{Fig:2} also indicates that the performance of the ADMM algorithm is slightly better than that of the MM algorithms (MM-SDR and MM-ADMM) under both constraints (but requires a longer running time). In addition, the relative entropy of the waveforms synthesized under the PAPR constraint is lower than that under the energy constraint. This is because that the feasibility region corresponding to the PAPR constraint is smaller than that corresponding to the energy constraint.

\begin{figure}[!htbp]
\centering
{{{\includegraphics[width = 0.4\textwidth]{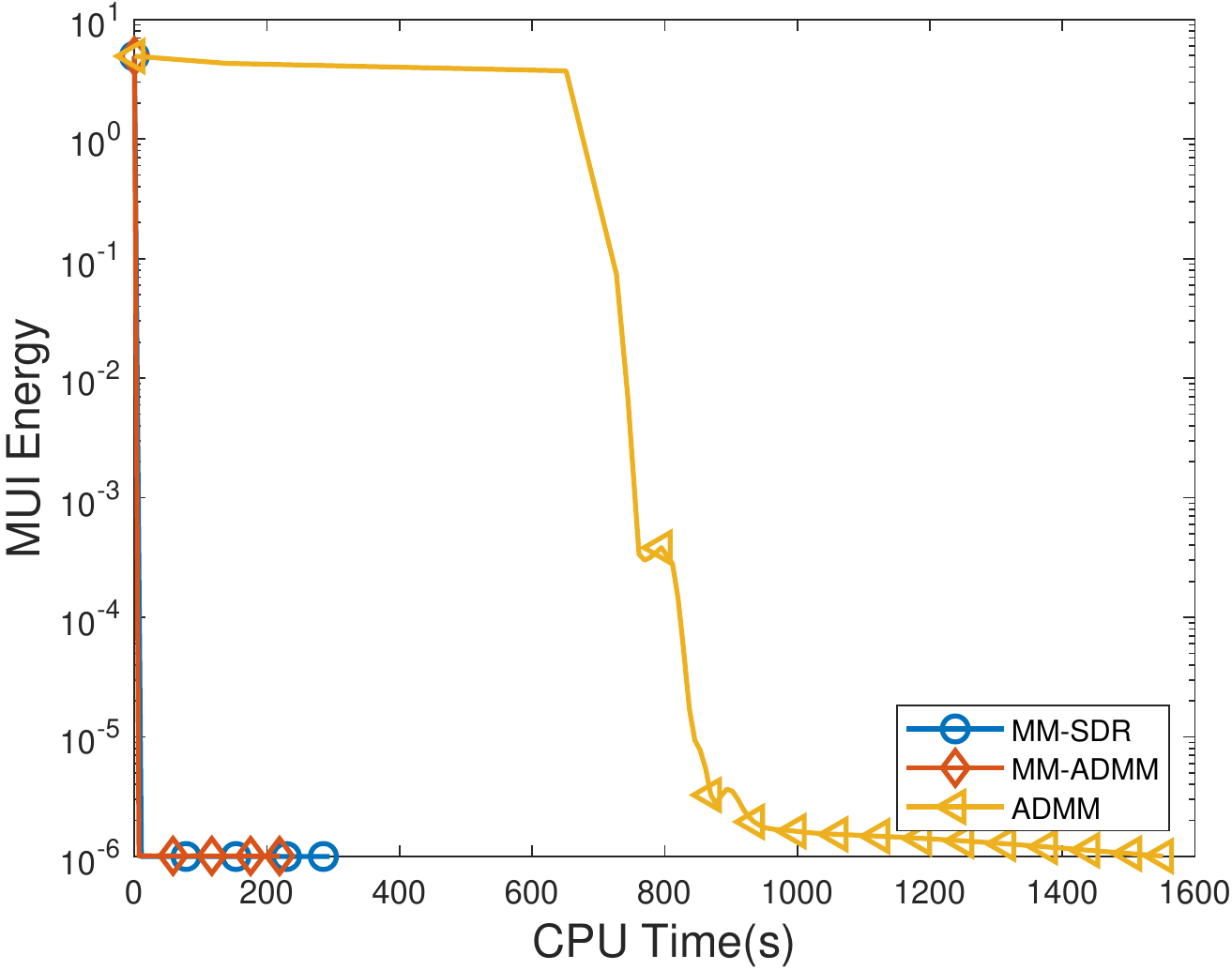}}} }
\caption{MUI energy of the synthesized waveforms versus the CPU time ($P_\textrm{t} = 1$). }
\label{Fig:3}
\end{figure}
\begin{figure*}[!htbp]
\centering
\centering
{\subfigure[]{{\includegraphics[width = 0.4\textwidth]{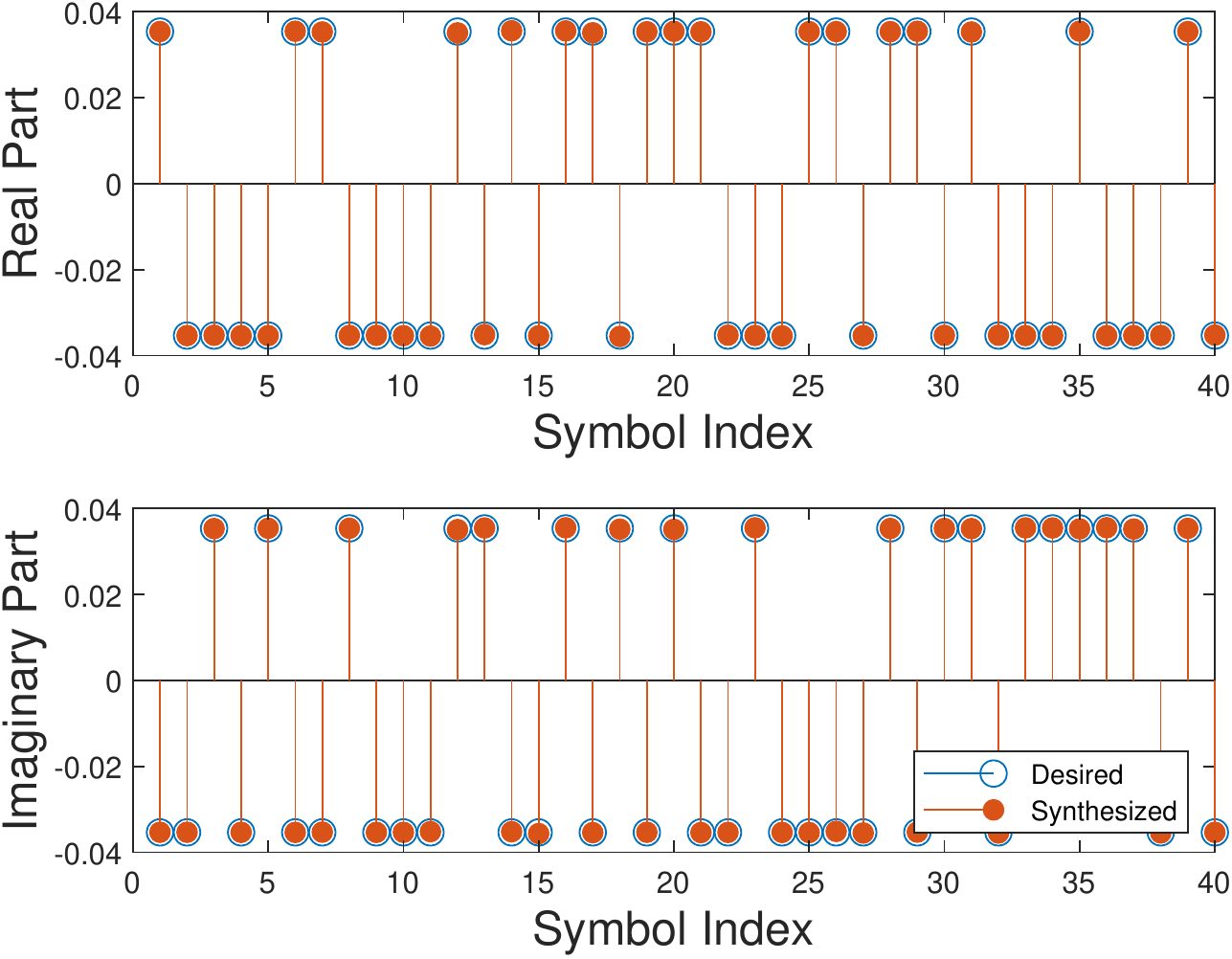}} \label{Fig:4a}} }
{\subfigure[]{{\includegraphics[width = 0.32\textwidth]{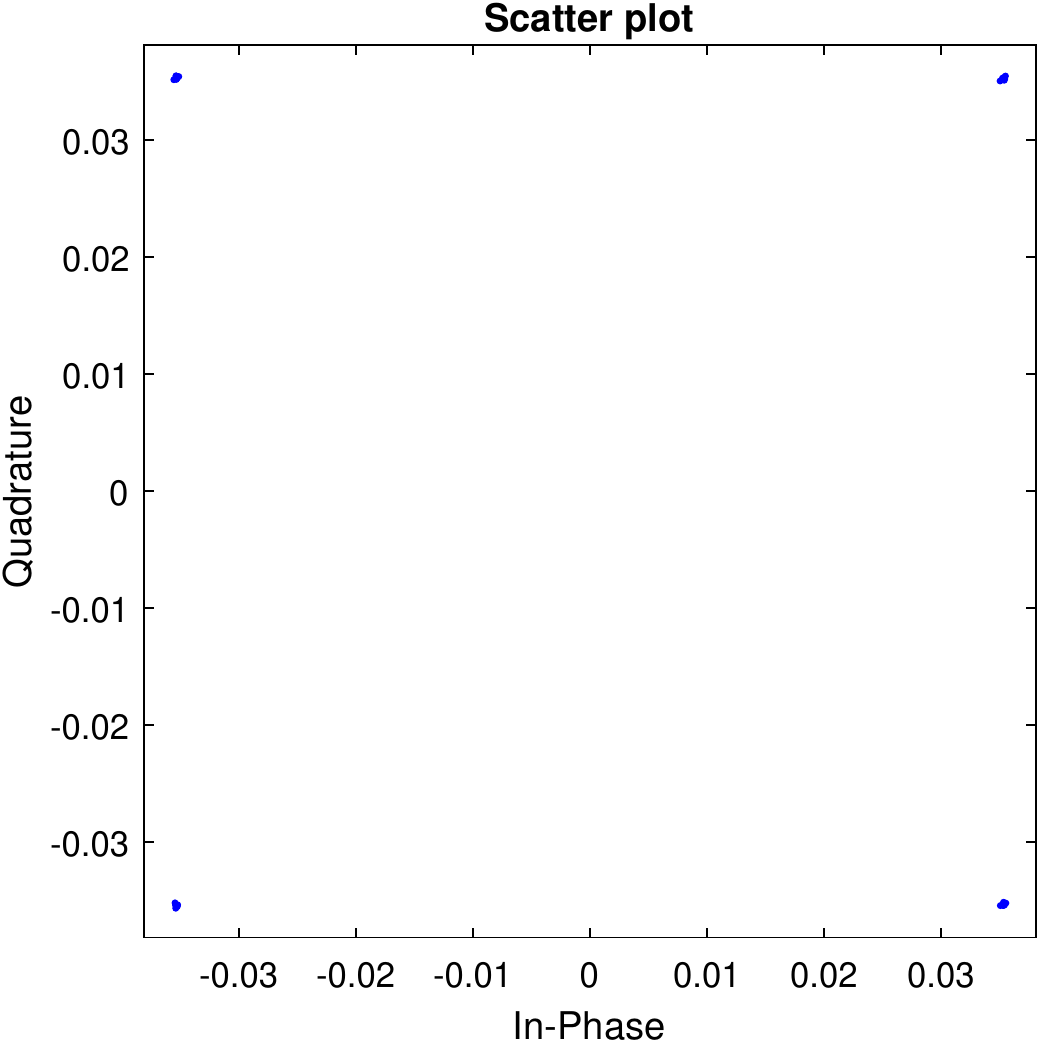}} \label{Fig:4b}} }
{\subfigure[]{{\includegraphics[width = 0.4\textwidth]{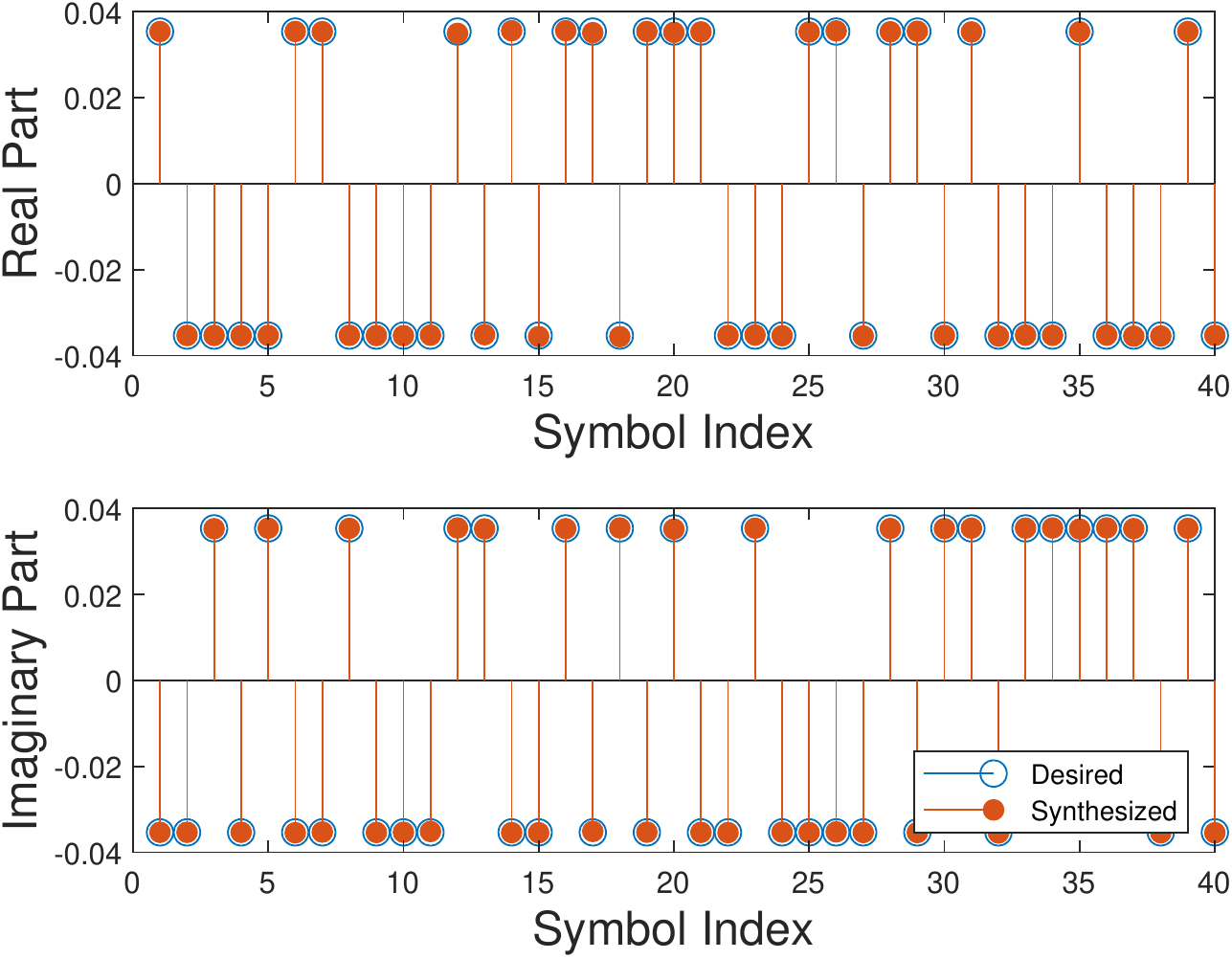}} \label{Fig:4c}} }
{\subfigure[]{{\includegraphics[width = 0.32\textwidth]{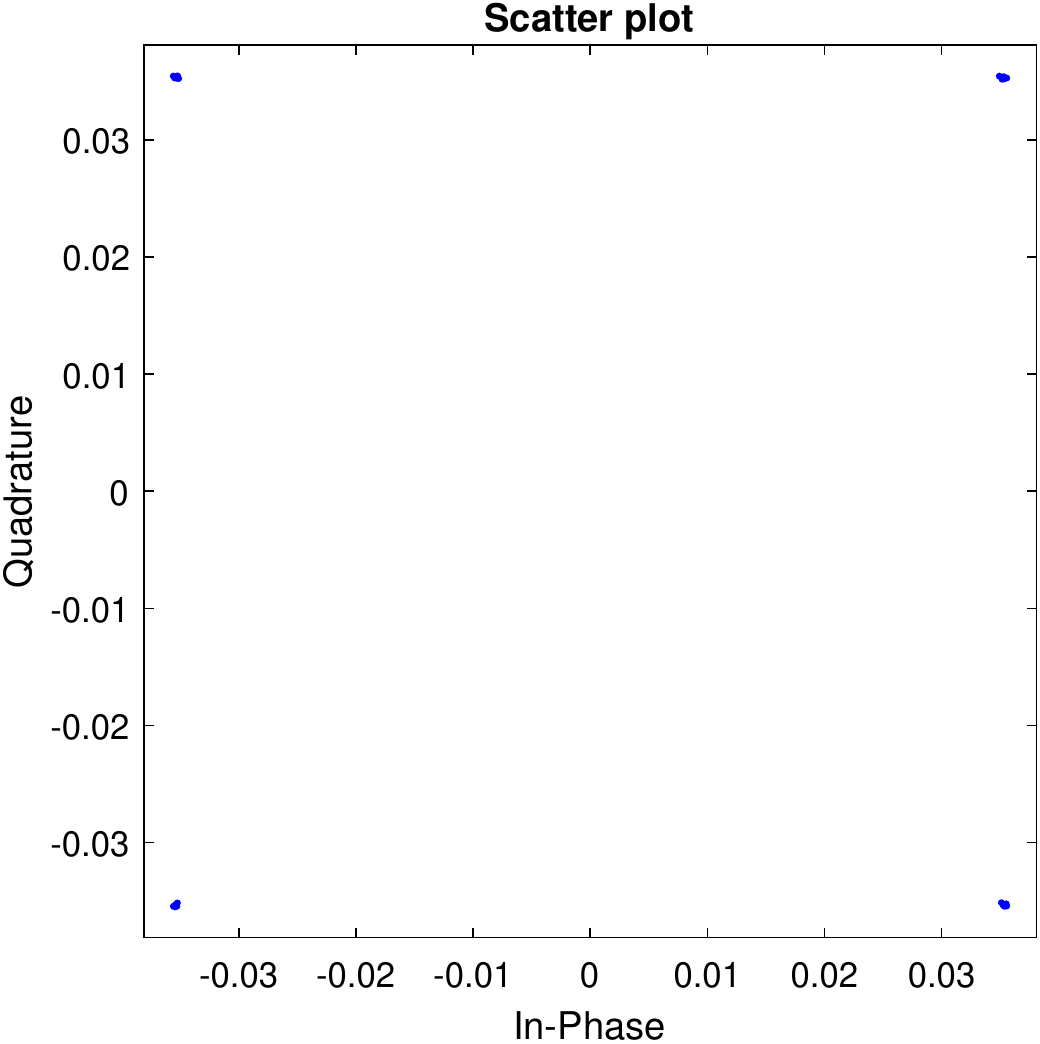}} \label{Fig:4d}} }
{\subfigure[]{{\includegraphics[width = 0.4\textwidth]{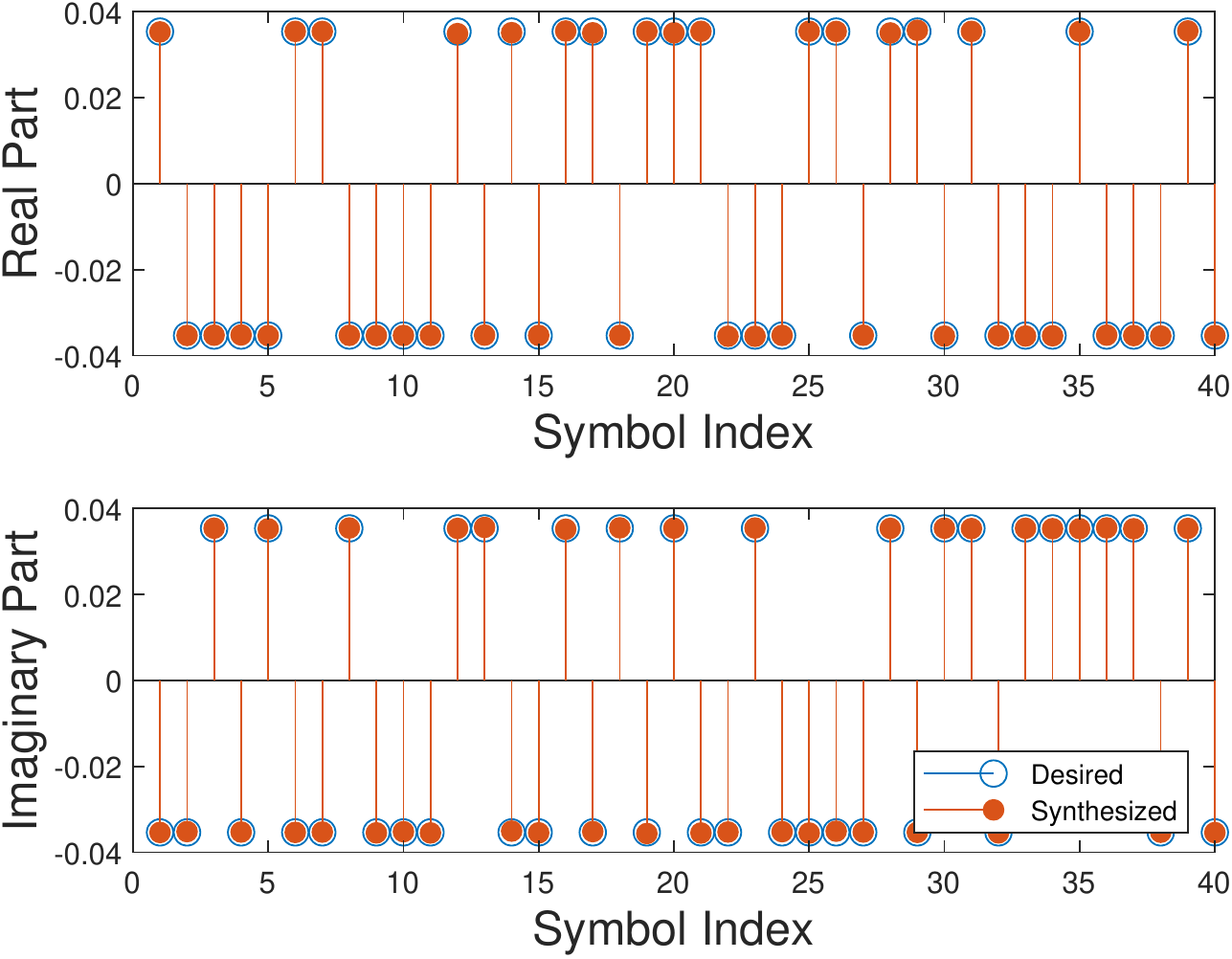}} \label{Fig:4e}} }
{\subfigure[]{{\includegraphics[width = 0.32\textwidth]{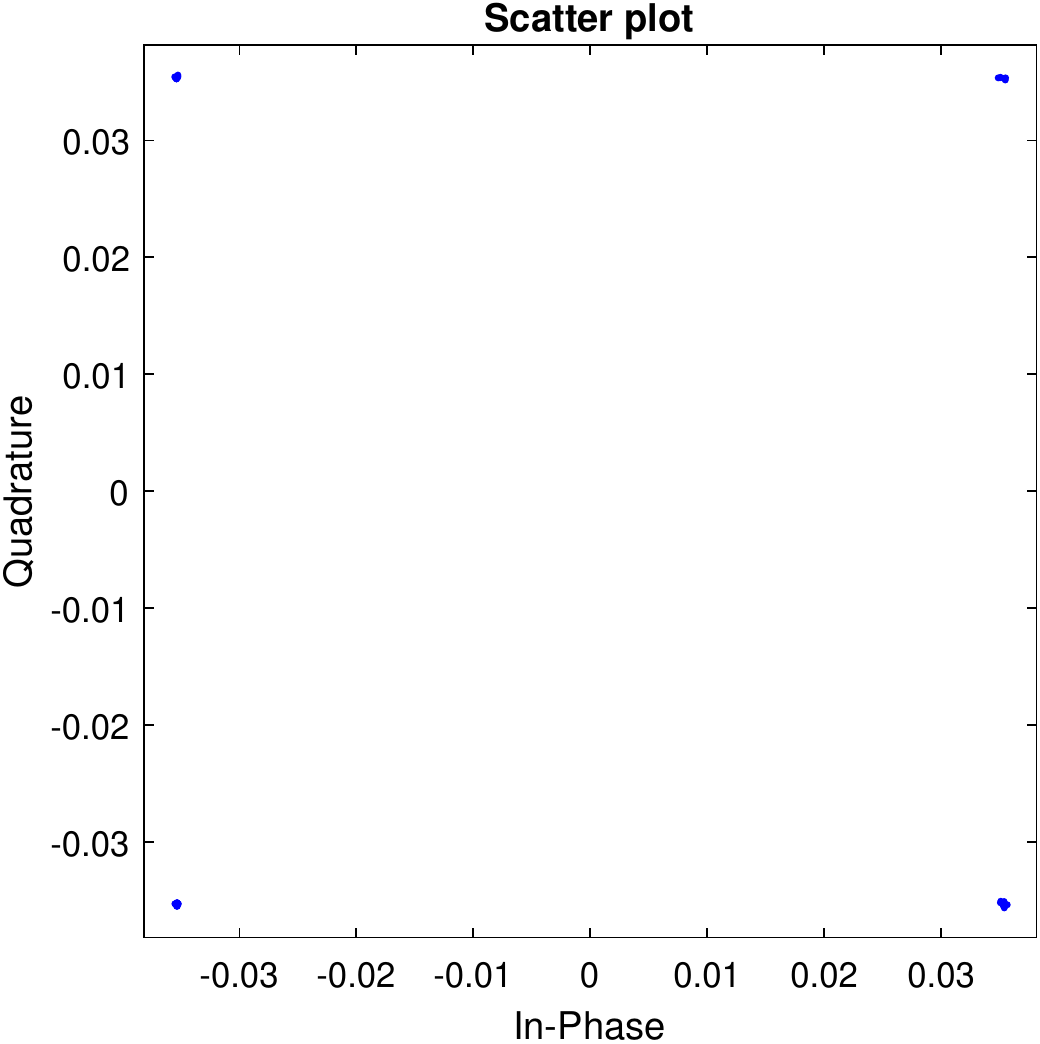}} \label{Fig:4f}} }
\caption{(a) Comparison of the communication signals synthesized by the MM-SDR algorithm with the desired ones. (b) The constellation diagram of the synthesized communication signals associated with  Fig. \ref{Fig:4a}. (c) Comparison of the communication signals synthesized by the MM-ADMM algorithm with the desired ones. (d) The constellation diagram of the synthesized communication signals associated with Fig. \ref{Fig:4c}. (e) Comparison of the communication signals synthesized by the ADMM algorithm with the desired ones. (f) The constellation diagram of the synthesized communication signals associated with Fig. \ref{Fig:4e}.}
\label{Fig:4}
\end{figure*}

To show that the synthesized waveforms support data communications, Fig. \ref{Fig:3} draws the synthesis errors of the communication signals associated with the waveforms in Fig. \ref{Fig:2} versus the CPU time. We can find that the synthesis errors of all the synthesized communication signals at convergence satisfy the MUI energy constraints, implying that the distortions of the synthesized communication signals are small and the quality of communication service can be ensured. To verify this claim, Fig. \ref{Fig:4a}, Fig. \ref{Fig:4c}, and Fig. \ref{Fig:4e} show the communication signals synthesized by the MM-SDR algorithm, the MM-ADMM algorithm, and the ADMM algorithm, respectively. Fig. \ref{Fig:4b}, Fig. \ref{Fig:4d}, and Fig. \ref{Fig:4f} show the associated constellation diagrams. We can see that all the synthesized communication signals have small matching errors and perfect constellation diagrams.

\begin{figure}[!htbp]
\centering
{\includegraphics[width = 0.4\textwidth]{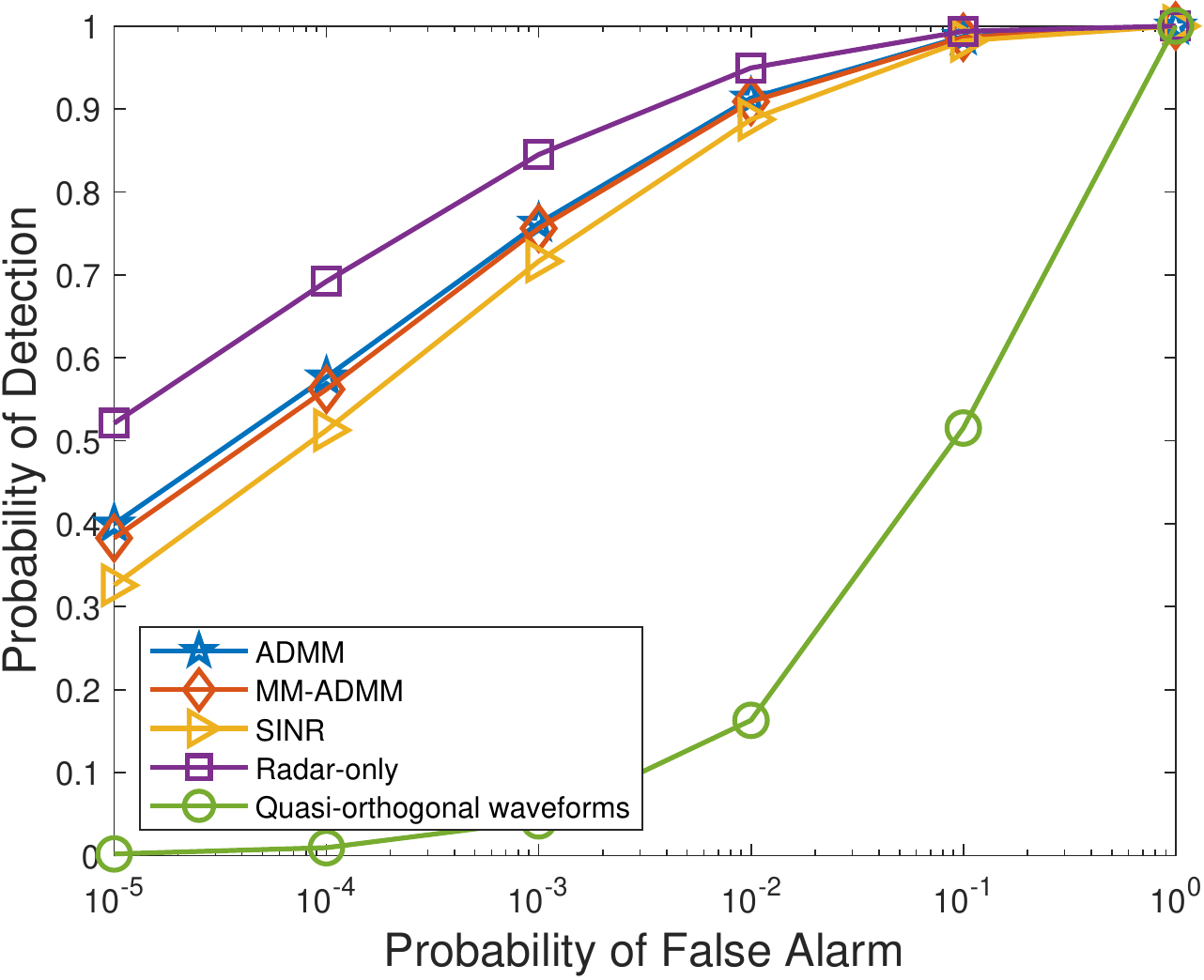}}
\caption{ROC of the synthesized waveforms ($P_\textrm{t} = 1$).}
\label{Fig:5}
\end{figure}

Next we analyze the receiver operating characteristic (ROC) associated with the quasi-orthogonal waveforms, the waveforms synthesized by the MM method, the waveforms synthesized by the ADMM algorithm, the waveforms designed based on maximizing SINR (the algorithm proposed in \cite{Wu2022dual-function} can be modified to synthesize the SINR-based waveforms.), and the radar-only waveforms (which is synthesized by removing the MUI energy constraint). To this purpose, we design an NP detector according to the hypothesis test established in \eqref{eq:hypothesistest}. The NP detector is given by:
\begin{align} \label{eq:NPdetector}
&\by^H(\bI - \bR_1^{-1})\by+2\Re[\by^H\bR_1^{-1}\widetilde{\bX}\bg_d]\mathop{\gtrless}\limits_{{{\cal H}_0}}^{{{\cal H}_1}}\gamma,
\end{align}
where $\gamma$ is the detection threshold. To design the waveforms, we consider the energy constraint and the transmit energy is $P_\textrm{t}=1$. Fig.~\ref{Fig:5} compares the ROC of these waveforms, where $10^7$ Monte Carlo trails are carried out the draw the curves.  In addition, the desired communication signals (i.e., $\bS$) are changed during the Monte Carlo trials as well.
We can see that the performance of the waveforms synthesized by the ADMM algorithm is slightly better than that synthesized by the MM algorithm, which is consistent with the results in Fig. \ref{Fig:2}. Since the synthesized waveforms support a dual-function, their performance is inferior to that of radar-only waveforms, but still better than that of the SINR based waveforms and quasi-orthogonal waveforms.

\begin{figure}[!htbp]
\centering
{\includegraphics[width = 0.4\textwidth]{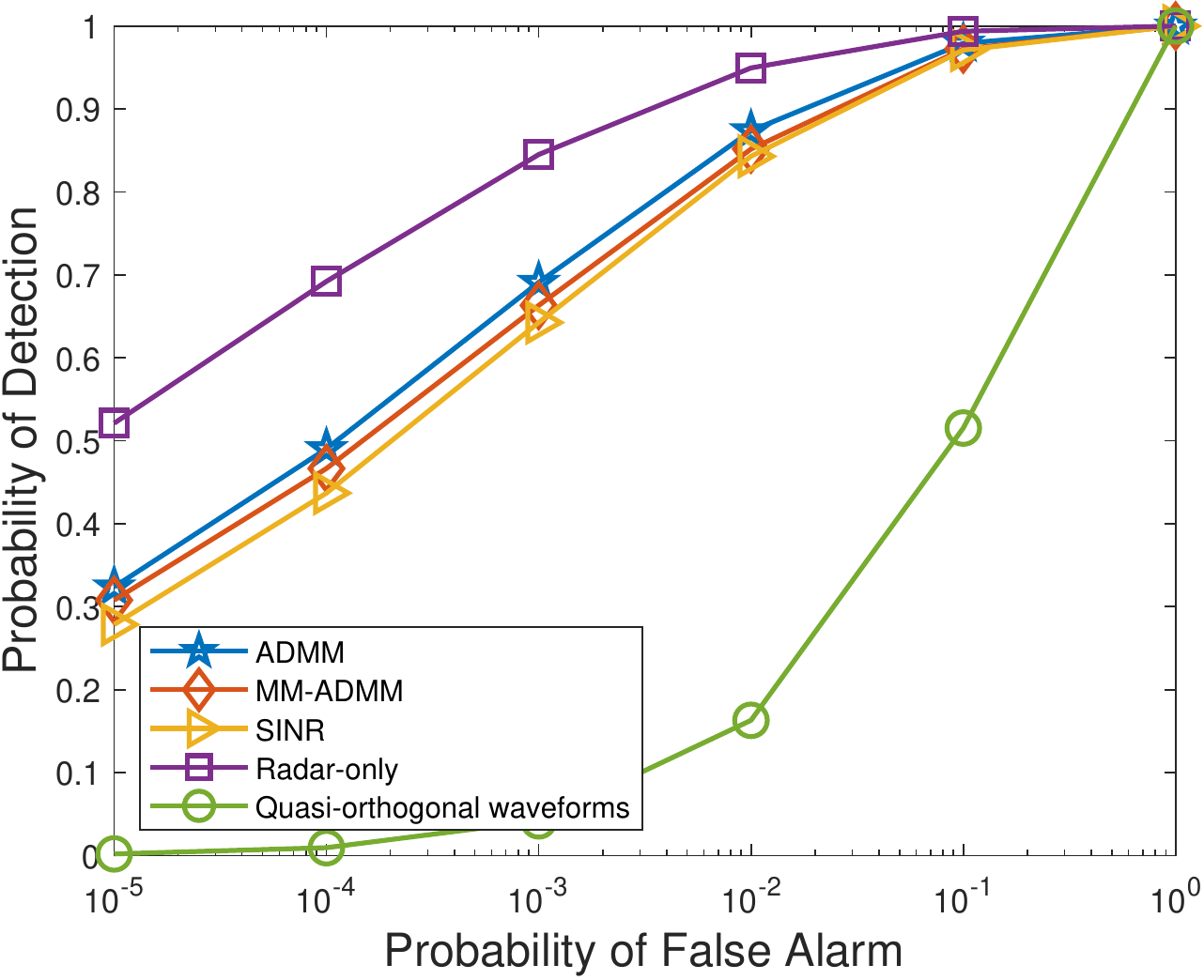}}
\caption{ROC averaged over a time-varying channel ($P_\textrm{t} = 1$).}
\label{Fig:6}
\end{figure}

To analyze the impact of a time-varying channel on the ROC performance of the waveforms, we compare the average ROC of these waveforms in Fig.~\ref{Fig:6}, where we consider 50 independent  realizations of $\bH$, and the other parameter settings are the same as those in Fig. \ref{Fig:5}. We can observe that the change of the channel matrix does not affect the ROC performance significantly, implying that the detection performance of the DFRC system is stable in a flat fading environment.

\begin{figure}[!htbp]
\centering
{\subfigure[]{{\includegraphics[width = 0.4\textwidth]{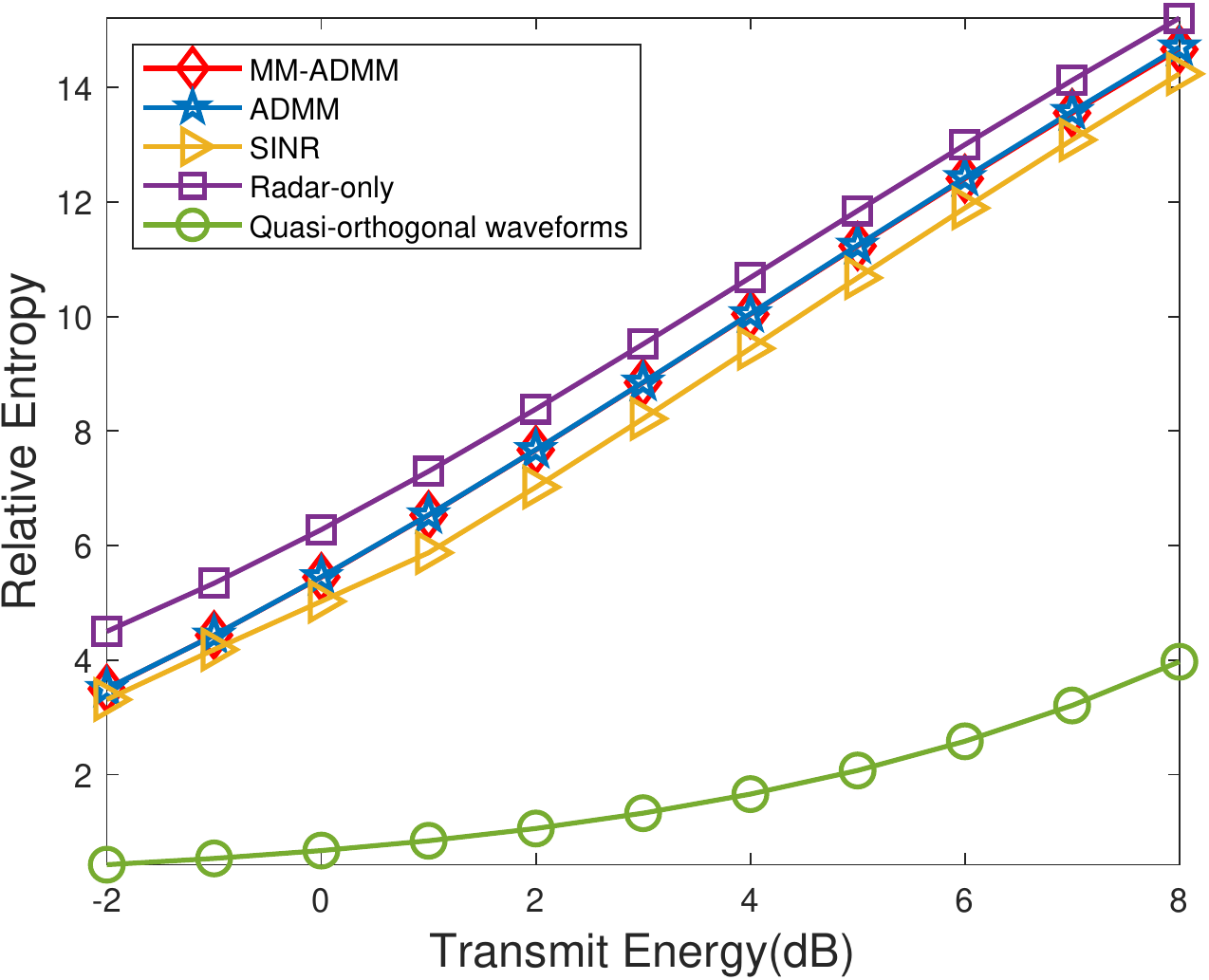}} \label{Fig:7a}} }
{\subfigure[]{{\includegraphics[width = 0.4\textwidth]{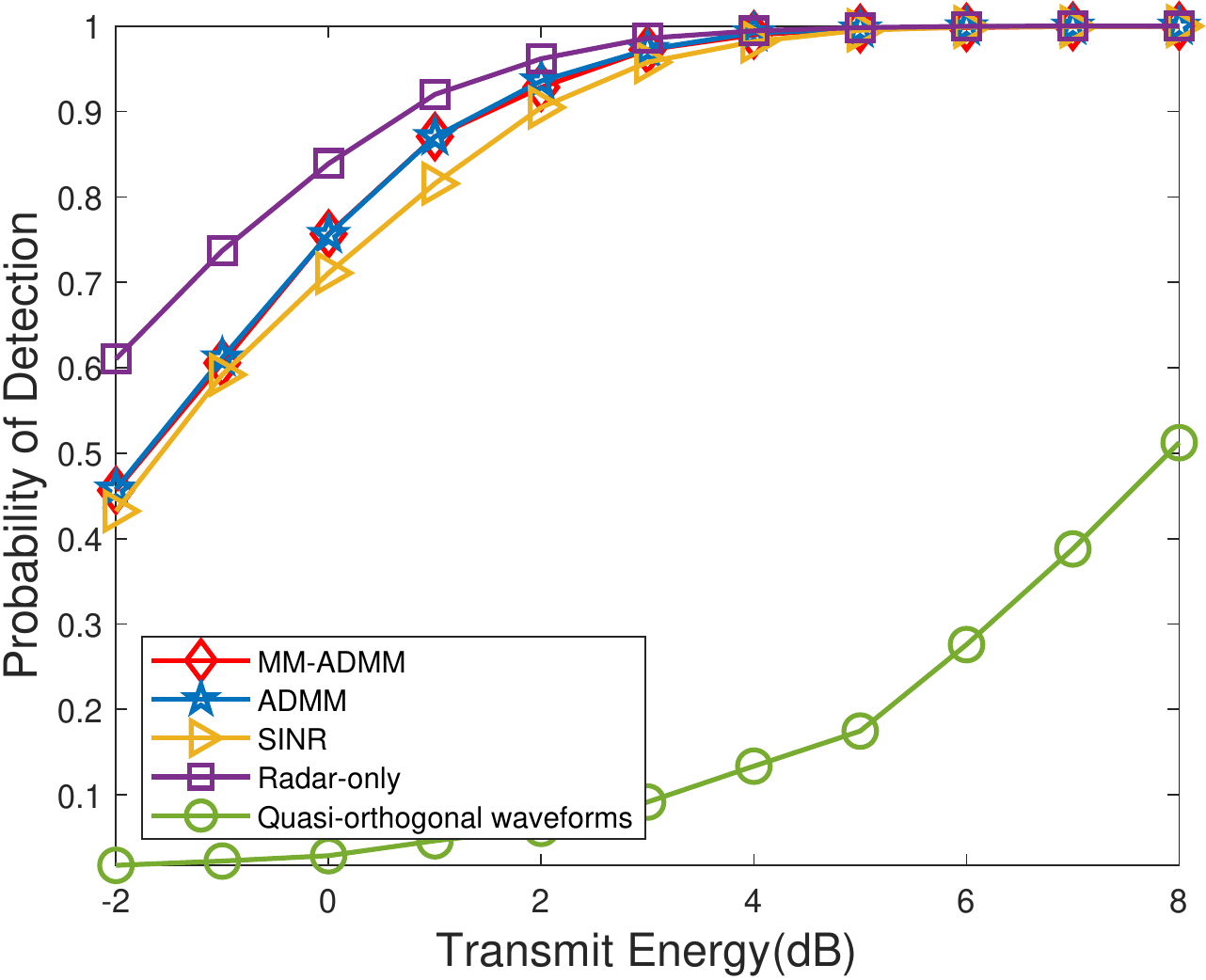}} \label{Fig:7b}} }
\caption{Performance of the synthesized waveforms. $P_{f_a}=10^{-3}$. (a) Relative entropy of the synthesized waveforms versus the transmit energy. (b) Detection performance corresponding to Fig. \ref{Fig:7a}.}
\label{Fig:7}
\end{figure}

Fig. \ref{Fig:7a} shows the relative entropy  of the five kinds of waveforms versus the transmit energy.
Fig. \ref{Fig:7b} presents the detection probabilities associated with the waveforms in Fig. \ref{Fig:7a}. The false alarm probability is fixed to be $P_{fa}=10^{-3}$. Again, we can see that the performance of the waveforms synthesized by the ADMM algorithm is slightly better than that synthesized by the MM algorithm in terms of relative entropy and the probability of detection, which is consistent with the results in Fig. \ref{Fig:5}. The performance of both the proposed algorithms are worse than the radar-only waveforms, but still better than the SINR-based waveforms and quasi-orthogonal waveforms.

\begin{figure}[!htbp]
\centering
{\includegraphics[width = 0.4\textwidth]{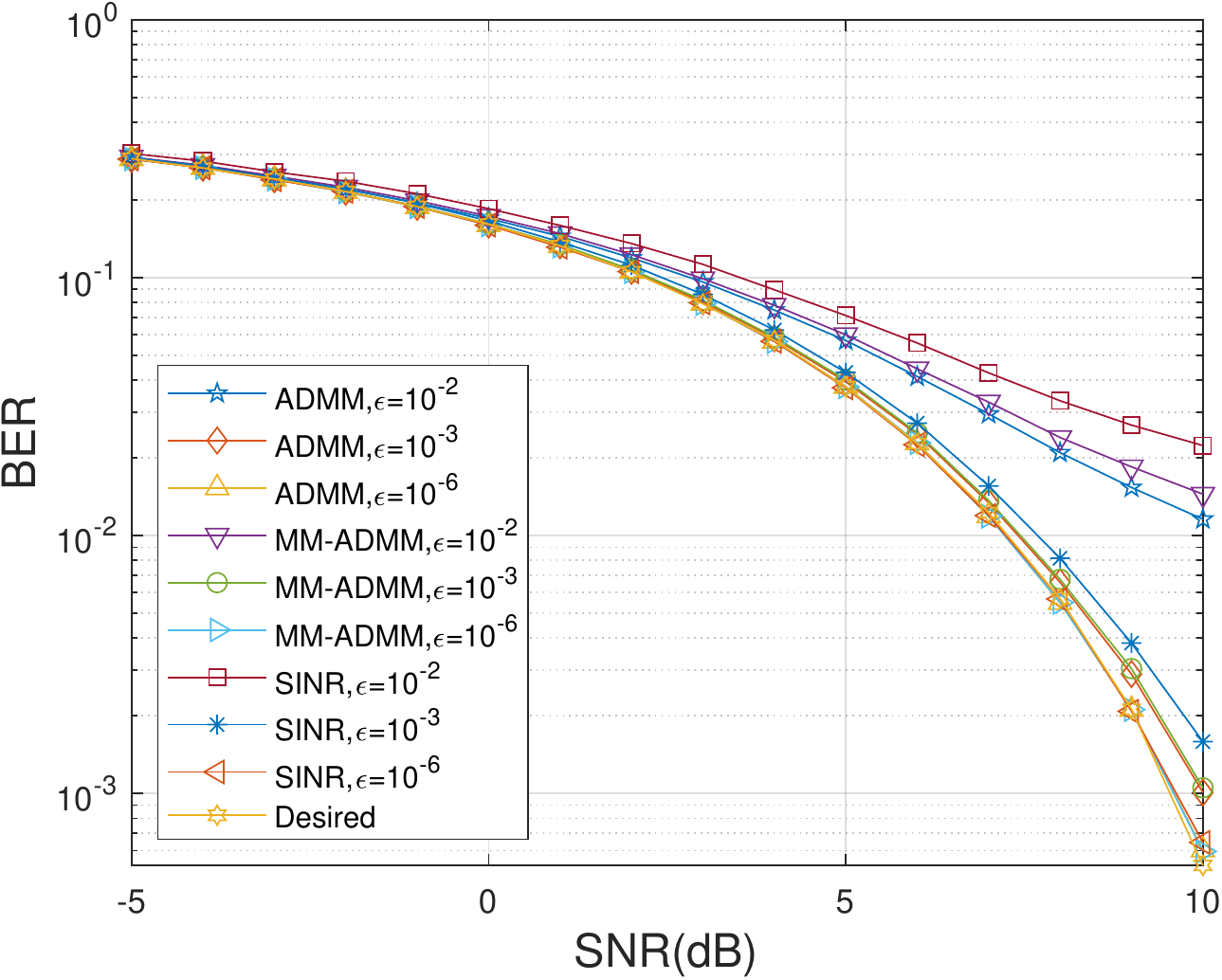}}
\caption{BER versus SNR for different $\epsilon$ ($P_\textrm{t} = 1$).}
\label{Fig:8}
\end{figure}

\begin{figure}[!htbp]
\centering
{\includegraphics[width = 0.4\textwidth]{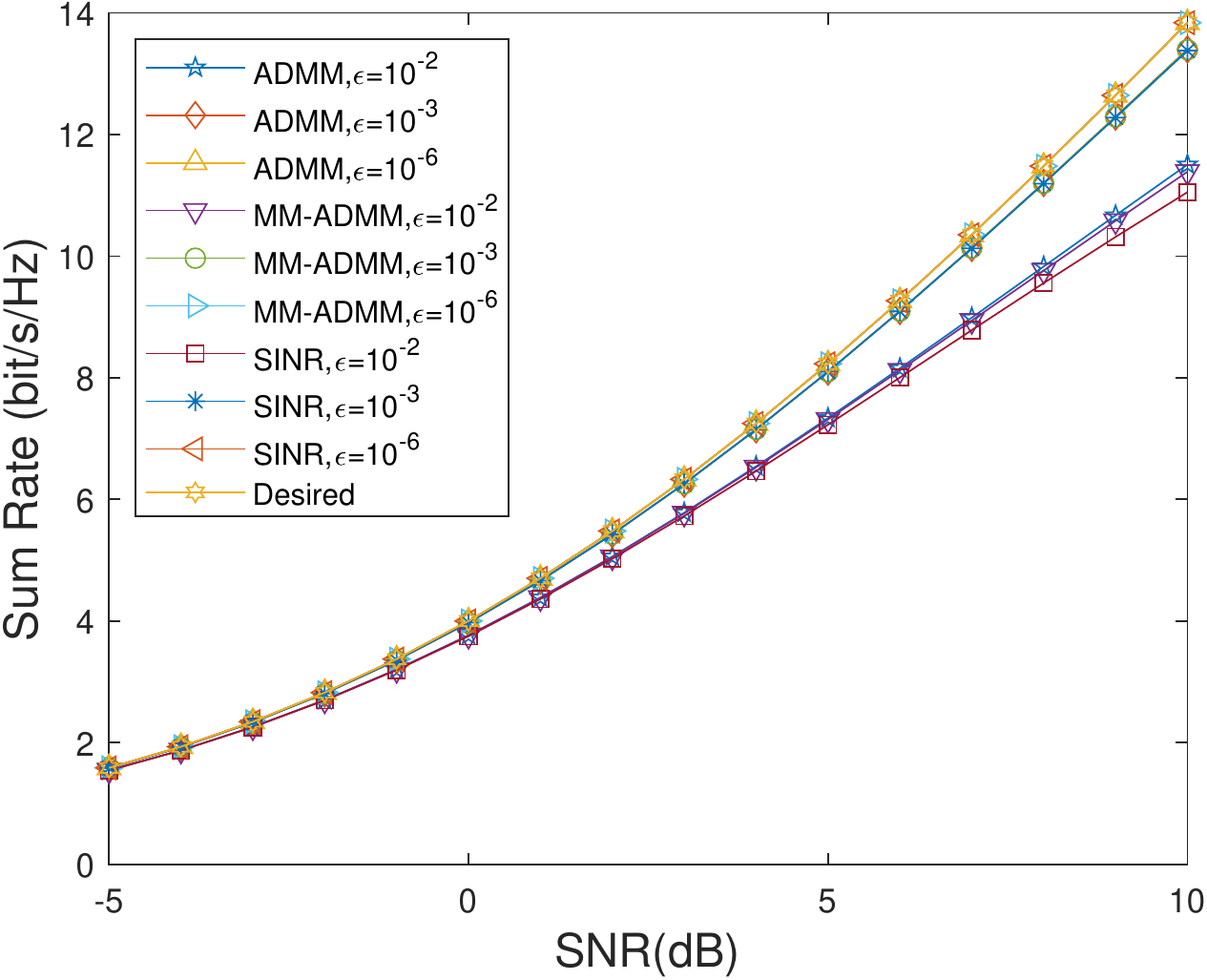}}
\caption{Achievable sum-rate versus SNR for different $\epsilon$ ($P_\textrm{t} = 1$).}
\label{Fig:9}
\end{figure}

Next we assess the performance of the synthesized communication signals. Fig. \ref{Fig:8} shows the bit error rate (BER) of the communication signals synthesized by the proposed algorithms versus the SNR, where $10^6$ independent Monte Carlo trails are conducted to draw the curves, the SNR is defined as  $$\textsf{SNR}={\mathbb{E}\{|s_{m,l}|^2\}}/{\sigma^2_{z,m}},$$
$s_{m,l}$ and $\sigma^2_{z,m}$ are the $l$th symbol of $\bs_m$ and the noise power in the $m$th communication receiver, respectively, and we have assumed that  the noise power for each user is equal.  We can find that the BER performance improves as the MUI energy decreases. In addition, the BER performance of the waveforms synthesized by the proposed algorithms is better than that of the SINR-based waveforms. Moreover, for the case of $\epsilon = 10^{-6}$, the performance of the synthesized signals is very close to that of the desired signals. Fig. \ref{Fig:9} shows the achievable sum rate of the communication signals synthesized by the presented algorithms versus the SNR. Again, the results show that a smaller MUI energy results in a better communication performance.  

\begin{figure}[!htbp]
\centering
{{\includegraphics[width = 0.4\textwidth]{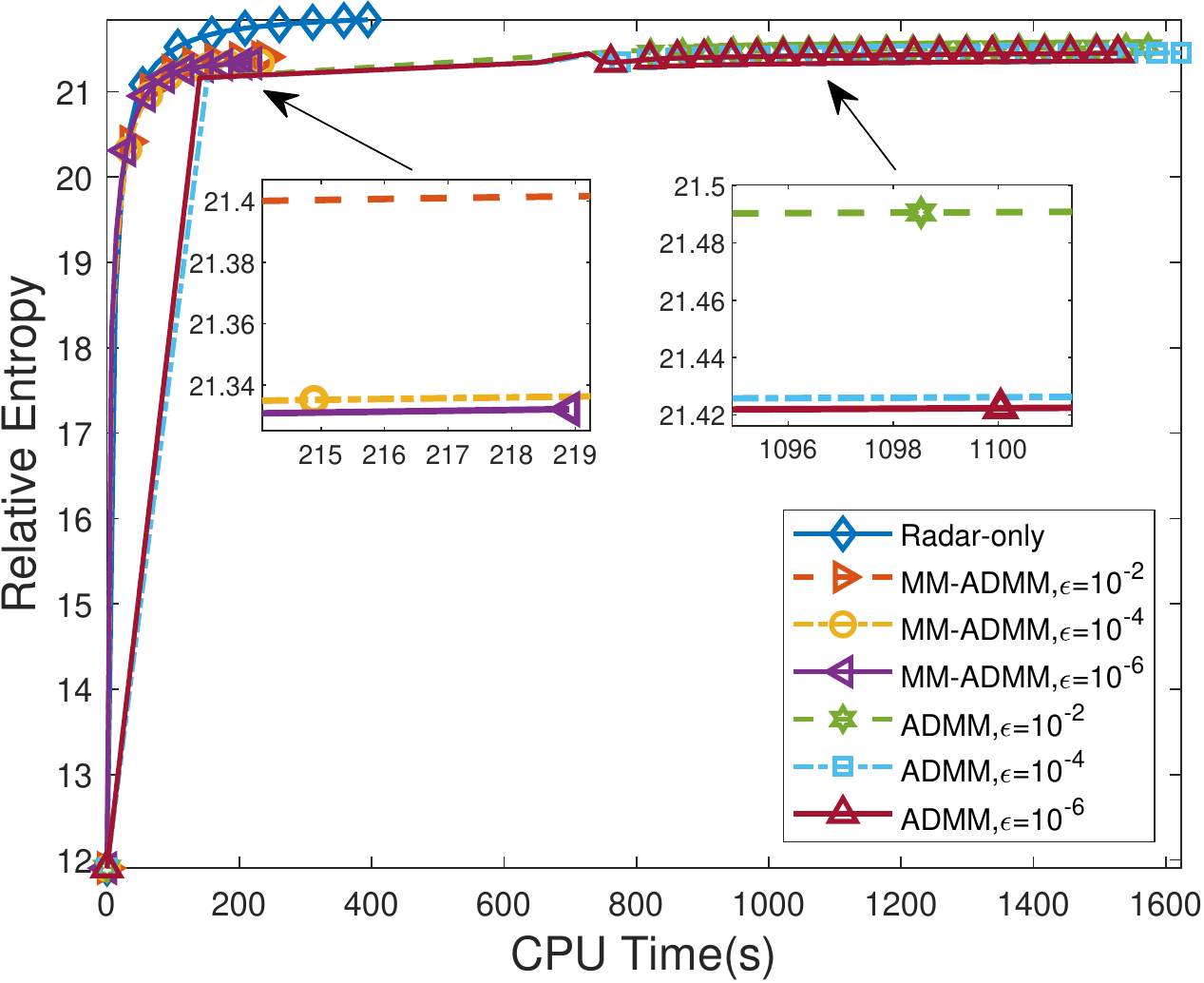}}}
\caption{Relative entropy for different $\epsilon$ ($P_\textrm{t} = 1$).}
\label{Fig:10}
\end{figure}

Fig. \ref{Fig:10} compares the relative entropy of the energy-constrained waveforms synthesized by the proposed algorithms for different MUI energy. As shown in the figure, even if we enforce a stringent MUI energy constraint on the communication signals, the detection performance of the synthesized waveforms only degrades slightly. 
\begin{figure}[!htbp]
\centering
{{\includegraphics[width = 0.4\textwidth]{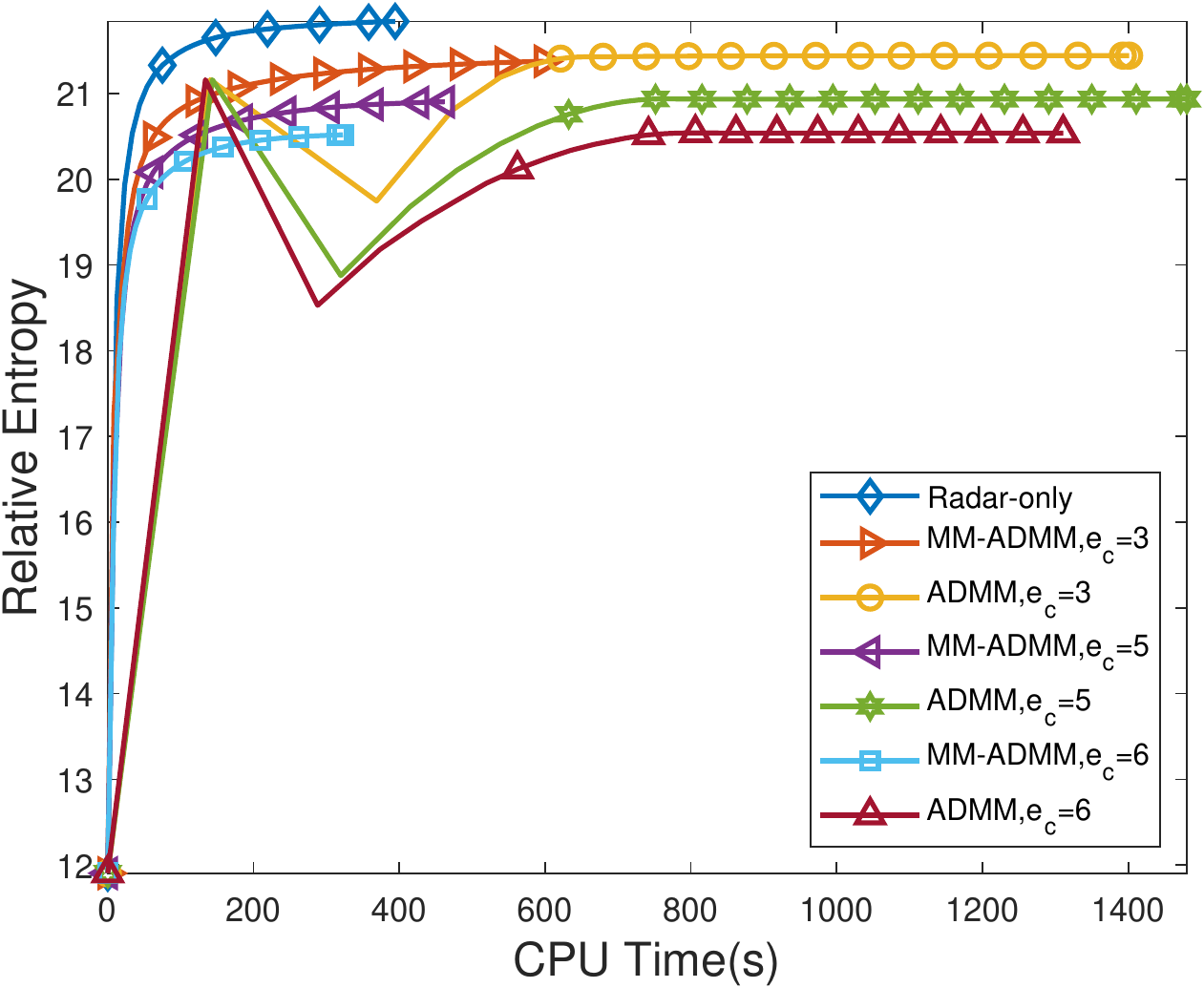}}}
\caption{Relative entropy for different $e_\textrm{c}$ ($P_\textrm{t} = 1$).}
\label{Fig:11}
\end{figure}

\begin{figure}[!htbp]
\centering
{{\includegraphics[width = 0.4\textwidth]{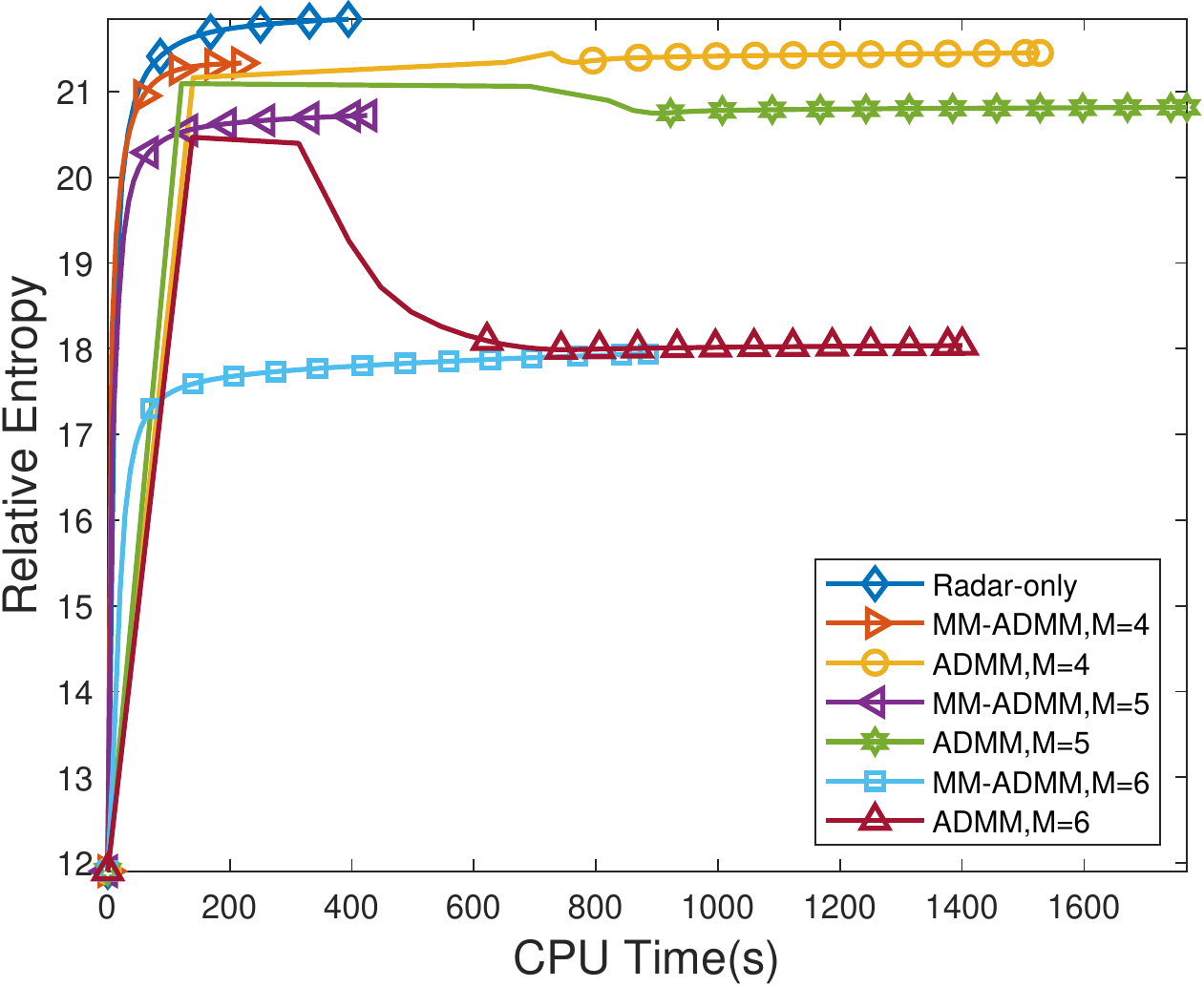}}}
\caption{Relative entropy for different $M$ ($P_\textrm{t} = 1$).}
\label{Fig:12}
\end{figure}

Fig. \ref{Fig:11} shows the relative entropy of the energy-constrained waveforms synthesized by the proposed algorithms for different energy of the communication signals (i.e., $e_\textrm{c}$). The results show that the relative entropy of the synthesized waveforms decreases with $e_\textrm{c}$. This is because that the MIMO DFRC system has to allocate more energy toward the communication users, which degrades the target detection performance. Finally, we analyze the impact of the number of users on the relative entropy of the waveforms synthesized by the proposed algorithms in Fig. \ref{Fig:12}. It can be found that as the number of users grows, the relative entropy of the synthesized waveforms becomes smaller, implying that the detection performance worsens.

\section{Conclusion} \label{Sec:conclusion}
This paper presented waveform design algorithms for MIMO DFRC systems. The goal was to improve the detection performance of a Rician target and ensure the quality of service for communications. We used the relative entropy as the design metric, and imposed a constraint on the communication MUI energy. To tackle the non-convex optimization problem, we proposed an MM-based approach and an ADMM algorithm. Simulation results showed that the waveforms synthesized by the presented algorithms could obtain better target detection performance and ensure the quality of communication service. In addition, the presented algorithm can be extended to design PAPR-constrained waveforms.

Possible topics for future research include the development of computationally efficient algorithms, which would enable the change of waveforms from frame to frame; the design of waveforms for DFRC systems in the presence of (signal-dependent) clutter; robust waveform design in the presence of prior knowledge mismatch (see, e.g., \cite{Naghsh2017Information} for a discussion on this topic).

\bibliographystyle{IEEEtran}
\bibliography{Rician_BIB}

\begin{thebibliography}{10}
\providecommand{\url}[1]{#1}
\csname url@samestyle\endcsname
\providecommand{\newblock}{\relax}
\providecommand{\bibinfo}[2]{#2}
\providecommand{\BIBentrySTDinterwordspacing}{\spaceskip=0pt\relax}
\providecommand{\BIBentryALTinterwordstretchfactor}{4}
\providecommand{\BIBentryALTinterwordspacing}{\spaceskip=\fontdimen2\font plus
\BIBentryALTinterwordstretchfactor\fontdimen3\font minus
  \fontdimen4\font\relax}
\providecommand{\BIBforeignlanguage}[2]{{%
\expandafter\ifx\csname l@#1\endcsname\relax
\typeout{** WARNING: IEEEtran.bst: No hyphenation pattern has been}%
\typeout{** loaded for the language `#1'. Using the pattern for}%
\typeout{** the default language instead.}%
\else
\language=\csname l@#1\endcsname
\fi
#2}}
\providecommand{\BIBdecl}{\relax}
\BIBdecl

\bibitem{Griffiths2013Challenge}
H.~Griffiths, S.~Blunt, L.~Cohen, and L.~Savy, ``Challenge problems in spectrum
  engineering and waveform diversity,'' in \emph{IEEE Radar Conference
  (RadarCon13)}, 2013, Conference Proceedings, pp. 1--5.

\bibitem{Griffiths2015Spectrum}
H.~Griffiths, L.~Cohen, S.~Watts, E.~Mokole, C.~Baker, M.~Wicks, and S.~Blunt,
  ``Radar spectrum engineering and management: Technical and regulatory
  issues,'' \emph{Proceedings of the IEEE}, vol. 103, no.~1, pp. 85--102, 2015.

\bibitem{Rowe2014SHAPE}
W.~Rowe, P.~Stoica, and J.~Li, ``Spectrally constrained waveform design,''
  \emph{IEEE Signal Processing Magazine}, vol.~31, no.~3, pp. 157--162, 2014.

\bibitem{Tang2019shape}
B.~Tang and J.~Liang, ``Efficient algorithms for synthesizing probing waveforms
  with desired spectral shapes,'' \emph{IEEE Transactions on Aerospace and
  Electronic Systems}, vol.~55, no.~3, pp. 1174--1189, 2019.

\bibitem{Aubry2016Optimization}
A.~Aubry, V.~Carotenuto, A.~De.~Maio, A.~Farina, and L.~Pallotta,
  ``Optimization theory-based radar waveform design for spectrally dense
  environments,'' \emph{IEEE Aerospace and Electronic Systems Magazine},
  vol.~31, no.~12, pp. 14--25, 2016.

\bibitem{Aubry2016Multiple}
A.~Aubry, V.~Carotenuto, and A.~De~Maio, ``Forcing multiple spectral
  compatibility constraints in radar waveforms,'' \emph{IEEE Signal Processing
  Letters}, vol.~23, no.~4, pp. 483--487, 2016.

\bibitem{Aubry2020MultiSpectrally}
A.~Aubry, A.~De.~Maio, M.~A. Govoni, and L.~Martino, ``On the design of
  multi-spectrally constrained constant modulus radar signals,'' \emph{IEEE
  Transactions on Signal Processing}, vol.~68, pp. 2231--2243, 2020.

\bibitem{Tang2018ADMM}
B.~Tang, J.~Li, and J.~Liang, ``Alternating direction method of multipliers for
  radar waveform design in spectrally crowded environments,'' \emph{Signal
  Processing}, vol. 142, pp. 398--402, 2018.

\bibitem{Li2017Joint}
B.~Li and A.~P. Petropulu, ``Joint transmit designs for coexistence of {MIMO}
  wireless communications and sparse sensing radars in clutter,'' \emph{IEEE
  Transactions on Aerospace and Electronic Systems}, vol.~53, no.~6, pp.
  2846--2864, 2017.

\bibitem{Qian2018Transmit}
J.~Qian, Z.~He, N.~Huang, and B.~Li, ``Transmit designs for spectral
  coexistence of {MIMO} radar and {MIMO} communication systems,'' \emph{IEEE
  Transactions on Circuits and Systems II: Express Briefs}, vol.~65, no.~12,
  pp. 2072--2076, 2018.

\bibitem{Qian2018Joint}
J.~Qian, M.~Lops, Z.~Le, X.~Wang, and Z.~He, ``Joint system design for
  coexistence of {MIMO} radar and {MIMO} communication,'' \emph{IEEE
  Transactions on Signal Processing}, vol.~66, no.~13, pp. 3504--3519, 2018.

\bibitem{Tavik2005RF}
G.~C. Tavik, C.~L. Hilterbrick, J.~B. Evins, J.~J. Alter, J.~G. Crnkovich,
  J.~W.~d. Graaf, W.~Habicht, G.~P. Hrin, S.~A. Lessin, D.~C. Wu, and S.~M.
  Hagewood, ``The advanced multifunction {RF} concept,'' \emph{IEEE
  Transactions on Microwave Theory and Techniques}, vol.~53, no.~3, pp.
  1009--1020, 2005.

\bibitem{Liu2020JRC}
F.~Liu, C.~Masouros, A.~P. Petropulu, H.~Griffiths, and L.~Hanzo, ``Joint radar
  and communication design: Applications, state-of-the-art, and the road
  ahead,'' \emph{IEEE Transactions on Communications}, vol.~68, no.~6, pp.
  3834--3862, 2020.

\bibitem{Zhang2021Overview}
J.~A. Zhang, F.~Liu, C.~Masouros, R.~W. Heath, Z.~Feng, L.~Zheng, and
  A.~Petropulu, ``An overview of signal processing techniques for joint
  communication and radar sensing,'' \emph{IEEE Journal of Selected Topics in
  Signal Processing}, vol.~15, no.~6, pp. 1295--1315, 2021.

\bibitem{Tang2022MFRF}
B.~Tang and P.~Stoica, ``{MIMO} multifunction {RF} systems: Detection
  performance and waveform design,'' \emph{IEEE Transactions on Signal
  Processing}, pp. 1--14, 2022.

\bibitem{Zhang2020Circulating}
Q.~Zhang, Y.~Zhou, L.~Zhang, Y.~Gu, and J.~Zhang, ``Circulating code array for
  a dual-function radar-communications system,'' \emph{IEEE Sensors Journal},
  vol.~20, no.~2, pp. 786--798, 2020.

\bibitem{Sanson2021Cooperative}
J.~B. Sanson, D.~Castanheira, A.~Gameiro, and P.~P. Monteiro, ``Cooperative
  method for distributed target tracking for {OFDM} radar with fusion of radar
  and communication information,'' \emph{IEEE Sensors Journal}, vol.~21,
  no.~14, pp. 15\,584--15\,597, 2021.

\bibitem{Bekar2021PSK-LFM}
M.~Bekar, C.~J. Baker, E.~G. Hoare, and M.~Gashinova, ``Joint {MIMO} radar and
  communication system using a {PSK}-{LFM} waveform with {TDM} and {CDM}
  approaches,'' \emph{IEEE Sensors Journal}, vol.~21, no.~5, pp. 6115--6124,
  2021.

\bibitem{Liu2022PAPR}
Y.~Liu, J.~Yi, X.~Wan, Y.~Rao, and J.~Shen, ``{PAPR} reduction of {OFDM}
  waveform in integrated passive radar and communication systems,'' \emph{IEEE
  Sensors Journal}, vol.~22, no.~17, pp. 17\,307--17\,317, 2022.

\bibitem{Liu2019Robust}
Y.~Liu, G.~Liao, and Z.~Yang, ``Robust {OFDM} integrated radar and
  communications waveform design based on information theory,'' \emph{Signal
  Processing}, vol. 162, pp. 317--329, 2019.

\bibitem{Liu2020OFDM}
Y.~Liu, G.~Liao, Y.~Chen, J.~Xu, and Y.~Yin, ``Super-resolution range and
  velocity estimations with {OFDM} integrated radar and communications
  waveform,'' \emph{IEEE Transactions on Vehicular Technology}, vol.~69,
  no.~10, pp. 11\,659--11\,672, 2020.

\bibitem{Liu2015MSK}
Z.~Liu, X.~Chen, X.~Wang, S.~Xu, and F.~Yuan, ``Communication analysis of
  integrated waveform based on {LFM} and {MSK},'' in \emph{IET International
  Radar Conference}, 2015, Conference Proceedings, pp. 1--5.

\bibitem{LiMIMObook2008}
J.~Li and P.~Stoica, \emph{{MIMO} Radar Signal Processing}.\hskip 1em plus
  0.5em minus 0.4em\relax Hoboken, NJ, USA: Wiley, 2008.

\bibitem{Shi2020Spectrally}
S.~Shi, Z.~Wang, Z.~He, and Z.~Cheng, ``Spectrally compatible waveform design
  for {MIMO} radar with {ISL} and {PAPR} constraints,'' \emph{IEEE Sensors
  Journal}, vol.~20, no.~5, pp. 2368--2377, 2020.

\bibitem{Sayin2020Design}
A.~Sayin, E.~G. Hoare, and M.~Antoniou, ``Design and verification of reduced
  redundancy ultrasonic {MIMO} arrays using simulated annealing \& genetic
  algorithms,'' \emph{IEEE Sensors Journal}, vol.~20, no.~9, pp. 4968--4975,
  2020.

\bibitem{Nan2020Beampattern}
N.~Liu, Z.~Zhang, and L.~Zhang, ``Waveform analytic design method for transmit
  beampattern synthesis of circulating coded {MIMO} radar,'' \emph{IEEE Sensors
  Journal}, vol.~20, no.~3, pp. 1485--1498, 2020.

\bibitem{Cong2021Robust}
J.~Cong, X.~Wang, M.~Huang, and L.~Wan, ``Robust {DOA} estimation method for
  {MIMO} {Radar} via deep neural networks,'' \emph{IEEE Sensors Journal},
  vol.~21, no.~6, pp. 7498--7507, 2021.

\bibitem{Liu2018Toward}
F.~Liu, L.~Zhou, C.~Masouros, A.~Li, W.~Luo, and A.~Petropulu, ``Toward
  dual-functional radar-communication systems: Optimal waveform design,''
  \emph{IEEE Transactions on Signal Processing}, vol.~66, no.~16, pp.
  4264--4279, 2018.

\bibitem{Tang2020Dualfunction}
B.~Tang, H.~Wang, L.~Qin, and L.~Li, ``Waveform design for dual-function {MIMO}
  radar-communication systems,'' in \emph{IEEE 11th Sensor Array and
  Multichannel Signal Processing Workshop (SAM)}, 2020, Conference Proceedings,
  pp. 1--5.

\bibitem{Tsinos2021Joint}
C.~G. Tsinos, A.~Arora, S.~Chatzinotas, and B.~Ottersten, ``Joint transmit
  waveform and receive filter design for dual-function radar-communication
  systems,'' \emph{IEEE Journal of Selected Topics in Signal Processing},
  vol.~15, no.~6, pp. 1378--1392, 2021.

\bibitem{Wu2022dual-function}
W.~Wu, B.~Tang, J.~Tang, and Y.~Hu, ``Waveform design for dual-function
  radar-communication systems in clutter,'' \emph{Journal of Radars}, vol.~11,
  no.~4, pp. 570--580, 2022.

\bibitem{Liu2022Cramer}
F.~Liu, Y.~F. Liu, A.~Li, C.~Masouros, and Y.~C. Eldar, ``Cramer-{R}ao {B}ound
  optimization for joint radar-communication beamforming,'' \emph{IEEE
  Transactions on Signal Processing}, vol.~70, pp. 240--253, 2022.

\bibitem{Liu2020Multiuser}
X.~Liu, T.~Huang, N.~Shlezinger, Y.~Liu, J.~Zhou, and Y.~C. Eldar, ``Joint
  transmit beamforming for multiuser {MIMO} communications and {MIMO} radar,''
  \emph{IEEE Transactions on Signal Processing}, vol.~68, pp. 3929--3944, 2020.

\bibitem{Liu2022Covariance}
X.~Liu, T.~Huang, and Y.~Liu, ``Transmit design for joint {MIMO} radar and
  multiuser communications with transmit covariance constraint,'' \emph{IEEE
  Journal on Selected Areas in Communications}, vol.~40, no.~6, pp. 1932--1950,
  2022.

\bibitem{Yuan2021Spatio-Temporal}
X.~Yuan, Z.~Feng, J.~A. Zhang, W.~Ni, R.~P. Liu, Z.~Wei, and C.~Xu,
  ``Spatio-temporal power optimization for {MIMO} joint communication and radio
  sensing systems with training overhead,'' \emph{IEEE Transactions on
  Vehicular Technology}, vol.~70, no.~1, pp. 514--528, 2021.

\bibitem{Tang2019Spectrally}
B.~Tang and J.~Li, ``Spectrally constrained {MIMO} radar waveform design based
  on mutual information,'' \emph{IEEE Transactions on Signal Processing},
  vol.~67, no.~3, pp. 821--834, 2019.

\bibitem{Tang2021RangeSpread}
B.~Tang and P.~Stoica, ``Information-theoretic waveform design for {MIMO} radar
  detection in range-spread clutter,'' \emph{Signal Processing}, vol. 182, p.
  107961, 2021.

\bibitem{Wang2022Rician}
X.~Wang, B.~Tang, and M.~Zhang, ``Optimisation of practically constrained
  waveforms for {Rician} target detection with multiple-input-multiple-output
  radar,'' \emph{IET Radar, Sonar \& Navigation}, vol.~16, no.~7, pp.
  1116--1130, 2022.

\bibitem{Tang2016Rician}
B.~Tang, J.~Tang, and Y.~Zhang, ``Design of multiple-input-multiple-output
  radar waveforms for {Rician} target detection,'' \emph{IET Radar, Sonar \&
  Navigation}, vol.~10, no.~9, pp. 1583--1593, 2016.

\bibitem{Kaystatistical1998}
S.~M. Kay, \emph{Fundamentals of statistical signal processing, vol. ii:
  detection theory}.\hskip 1em plus 0.5em minus 0.4em\relax Upper Saddle River,
  Newer Jersey: Prentice-Hall, 1998.

\bibitem{Naghsh2012}
M.~M. Naghsh and M.~Modarres-Hashemi, ``Exact theoretical performance analysis
  of optimum detector in statistical multi-input multi-output radars,''
  \emph{IET Radar, Sonar \& Navigation}, vol.~6, no.~2, pp. 99--111, 2012.

\bibitem{Naghsh2013Unified}
M.~M. Naghsh, M.~Modarres-Hashemi, S.~ShahbazPanahi, M.~Soltanalian, and
  P.~Stoica, ``Unified optimization framework for multi-static radar code
  design using information-theoretic criteria,'' \emph{IEEE Transactions on
  Signal Processing}, vol.~61, no.~21, pp. 5401--5416, 2013.

\bibitem{Naghsh2017Information}
M.~M. Naghsh, M.~Modarres-Hashemi, M.~A. Kerahroodi, and E.~H.~M. Alian, ``An
  information theoretic approach to robust constrained code design for {MIMO}
  radars,'' \emph{IEEE Transactions on Signal Processing}, vol.~65, no.~14, pp.
  3647--3661, 2017.

\bibitem{Tang2010mimo}
\BIBentryALTinterwordspacing
B.~Tang, J.~Tang, and Y.~Peng, ``{MIMO} radar waveform design in colored noise
  based on information theory,'' \emph{IEEE Transactions on Signal Processing},
  vol.~58, no.~9, pp. 4684--4697, 2010. [Online]. Available:
  \url{10.1109/TSP.2010.2050885}
\BIBentrySTDinterwordspacing

\bibitem{CoverInformationTheory1991}
T.~M. Cover and J.~Thomas, \emph{Elements of Information Theory}.\hskip 1em
  plus 0.5em minus 0.4em\relax New York, NY, USA: Wiley, 1991.

\bibitem{Mohammed2013Per-Antenna}
S.~K. Mohammed and E.~G. Larsson, ``Per-antenna constant envelope precoding for
  large multi-user {MIMO} systems,'' \emph{IEEE Transactions on
  Communications}, vol.~61, no.~3, pp. 1059--1071, 2013.

\bibitem{HornMatrix1990}
R.~A. Horn and C.~R. Jonson, \emph{Matrix Analysis}.\hskip 1em plus 0.5em minus
  0.4em\relax Cambridge, U.K: Cambridge Univ. Press, 1990.

\bibitem{TangMinorization2018}
B.~Tang, Y.~Zhang, and J.~Tang, ``An efficient minorization maximization
  approach for {MIMO} radar waveform optimization via relative entropy,''
  \emph{IEEE Transactions on Signal Processing}, vol.~66, no.~2, pp. 400--411,
  2018.

\bibitem{BoydConvex2004}
S.~Boyd and L.~Vandenberghe, \emph{Convex Optimization}.\hskip 1em plus 0.5em
  minus 0.4em\relax Cambridge, U.K: Cambridge Univ. Press, 2004.

\bibitem{HjorungnesDifferentiation2007}
A.~Hjorungnes and D.~Gesbert, ``Complex-valued matrix differentiation:
  Techniques and key results,'' \emph{IEEE Transactions on Signal Processing},
  vol.~55, no.~6, pp. 2740--2746, 2007.

\bibitem{Huang2007Complex}
Y.~Huang and S.~Zhang, ``Complex matrix decomposition and quadratic
  programming,'' \emph{Math. Oper. Res.}, vol.~32, no.~3, pp. 758--768, 2007.

\bibitem{Luo2010Semidefinite}
Z.-Q. Luo, W.-K. Ma, A.~Man-Cho~So, Y.~Ye, and S.~Zhang, ``Semidefinite
  relaxation of quadratic optimization problems,'' \emph{IEEE Signal Processing
  Magazine}, vol.~27, no.~3, pp. 20--34, 2010.

\bibitem{Stephen2008CVX}
M.~Grant and S.~Boyd, ``{CVX}: Matlab software for disciplined convex
  programming, version 1.21,'' \emph{Global Optimization}, pp. 155--210, 2008.

\bibitem{Ai2011Rank_one}
W.~Ai, Y.~Huang, and S.~Zhang, ``New results on {Hermitian} matrix rank-one
  decomposition,'' \emph{Math. Program.}, vol. 128, no. 1/2, pp. 253--283,
  2011.

\end{thebibliography}

\newpage

\section{Biography Section}
\vspace{-33pt}
\begin{IEEEbiography}[{\includegraphics[width=1in,height=1.25in,clip,keepaspectratio]{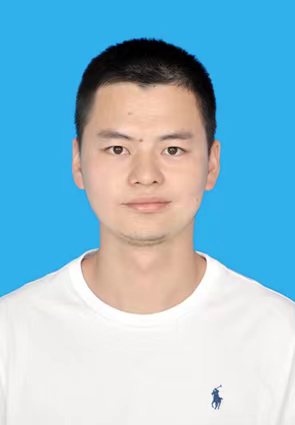}}]{Xuyang Wang} was born in Suihua, Heilongjiang Province, China, in 1992. He received the B.S. degree in electrical engineering from Electronic Engineering Institute, Hefei, China, in 2014 and the M.S. degree in electrical engineering from the College of Electronic Engineering, National University of Defense Technology, Hefei, China, in 2022. His research interests include radar signal processing and waveform design.
\end{IEEEbiography}

\vspace{-33pt}
\begin{IEEEbiography}[{\includegraphics[width=1in,height=1.25in,clip,keepaspectratio]{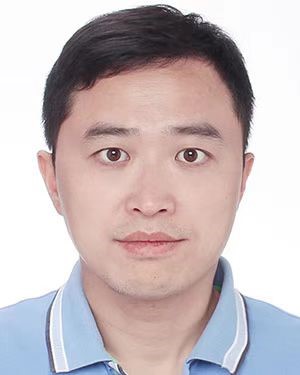}}]{Bo Tang} was born in Linchuan, Jiangxi Province, China, in 1985. He received the B.S. and Ph.D. degrees in electrical engineering from Tsinghua University, Beijing, China, in 2006 and 2011, respectively. From July 2011 to June 2017, he was with Electronic Engineering Institute, as a Lecturer. Since July 2017, he has been with the College of Electronic Engineering, National University of Defense Technology, Hefei, China, where he is currently a Professor. His research interests mainly include adaptive radar signal processing and radar waveform design. He was selected as the “Young Elite Scientists Sponsorship Program” by China Association for Science and Technology and sponsored by the Anhui Provincial Natural Science Foundation for Distinguished Young Scholars. He is currently an Associate Editor for the IEEE Transactions on Signal Processing.
\end{IEEEbiography}

\vspace{-33pt}
\begin{IEEEbiography}[{\includegraphics[width=1in,height=1.25in,clip,keepaspectratio]{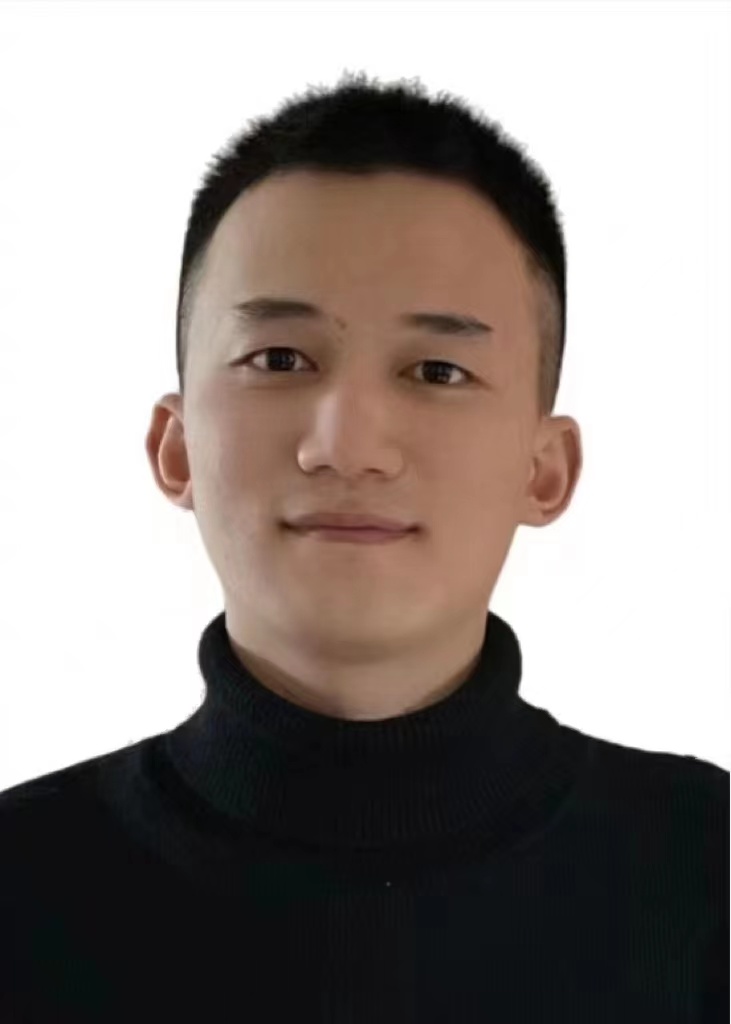}}]{Wenjun Wu} was born in Xuancheng, Anhui Province, China, in 1996. He received the B.S. degree in electrical engineering from the College of Electronic Engineering, National University of Defense Technology, Hefei, China, in 2018. He is currently pursuing the M.S. degree in electrical engineering with the College of Electronic Engineering, National University of Defense Technology. His research interests include radar signal processing and waveform design.
\end{IEEEbiography}

\vspace{-33pt}
\begin{IEEEbiography}[{\includegraphics[width=1in,height=1.25in,clip,keepaspectratio]{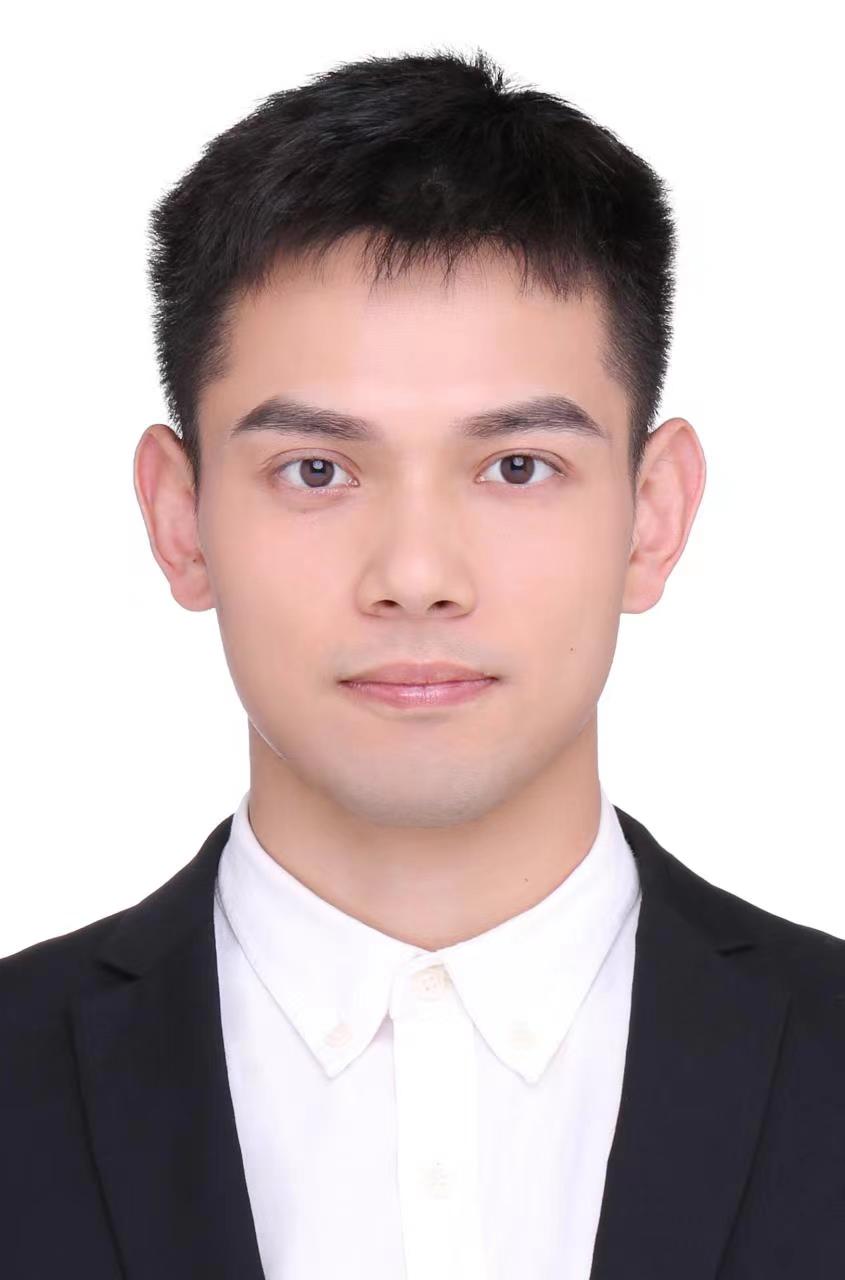}}]{Li Da} was born in Luoyang, Henan Province, China, in 1995. He received the B.S. degree in atmospheric sciences from Lanzhou University, Lanzhou, China, in 2020, the M.S. degree in electrical engineering from National University of Defense Technology, Hefei, China. He is currently working toward the Ph.D. degree in electrical engineering with the College of Electronic Engineering, National University of Defense Technology. His research interests mainly include radar waveform design and machine learning.
\end{IEEEbiography}

\end{document}